\documentclass[aps,prd,amsmath,11pt]{revtex4}                     %preprint,     
\usepackage{color,epsfig,graphicx,float,multirow,xspace,times}  
\definecolor{dblu}{rgb}{0.00,0.00,0.75}
\usepackage[colorlinks,urlcolor=dblu,linkcolor=dblu,citecolor=dblu]{hyperref} 
\usepackage[utf8]{inputenc} 
\def \babar {{\it BABAR} }
\begin{document} 

              \title{Low-mass enhancement of kaon pairs in $B^+\to\bar{D}^{(*)0} K^+\bar{K}^0$ and 
                        $B^0\to D^{(*)-}K^+\bar{K}^0$ decays}  
                    
\author{Wen-Fei Wang$^{1,2}$} \email{wfwang@ub.edu} 
\author{Li-Fei Yang$^{1}$} 
\author{Ai-Jun Ma$^{3}$}        
\author{\`Angels Ramos$^{2}$}  \email{ramos@fqa.ub.edu}

\affiliation{$^1$Institute of Theoretical Physics and State Key Laboratory of Quantum Optics and Quantum Optics Devices, 
                          Shanxi University, Taiyuan, Shanxi 030006, China\\
                    $^2$Departament de F\'{\i}sica Qu\`antica i Astrof\'{\i}sica and Institut de Ci\`encies del Cosmos (ICCUB), 
                         Facultat de F\'{\i}sica, Universitat de Barcelona, Mart\'i i Franqu\`es 1, 08028, Barcelona, Spain\\
	        $^3$School of Mathematics and Physics,  Nanjing Institute of Technology, Nanjing, Jiangsu 211167, China}  	         

\date{\today}

%XXXXXXXXXXXXXXXXXXXXXXXXXXXXXX%
\begin{abstract}
Very recently, the Belle~II Collaboration presented a measurement for the decays $B^+\to\bar{D}^{(*)0} K^+\bar{K}^0$ 
and $B^0\to D^{(*)-}K^+\bar{K}^0$, the bulk of observed $m(K^+ K_S^0)$ distributions showing low-mass structures 
in all four channels. In this work, we study the contributions of $\rho(770,1450)^+$, $a_2(1320)^+$ and $a_0(980,1450)^+$ 
resonances to these decay processes. The intermediate states $\rho(770,1450)^+$ are found to dominate the low-mass 
distribution of kaon pairs roughly contributing to half of the total branching fraction in each of the four decay channels. 
The contribution of the tensor $a_2(1320)^+$ meson is found to be negligible. Near the threshold of the kaon pair, the state 
$a_0(980)^+$ turns out to be much less important than expected, not being able to account for the enhancement of events 
in that energy region observed in the $B^+\to\bar{D}^{(*)0} K^+\bar{K}^0$ decays. Further studies both from the theoretical 
and experimental sides are needed to elucidate the role of the non-resonant contributions governing the formation of 
$K^+\bar{K}^0$ pairs near their threshold in these decay processes.
\end{abstract}
%XXXXXXXXXXXXXXXXXXXXXXXXXXXXXX%

%\pacs{13.20.He, 13.25.Hw, 13.30.Eg}

\maketitle

%%<<<<><><><><><><><><><><><><><><><><><><><><><><><><><><><><><><><><>>>>%%
\section{Introduction}%%%========SEC1
%%<<<<><><><><><><><><><><><><><><><><><><><><><><><><><><><><><><><><>>>>%%

Three-body hadronic $B$ meson decay processes are regularly interpreted in terms of the contribution of various resonant 
states.  The investigation of appropriate decay channels will help us to comprehend the properties and substructures 
of the related hadronic resonances involved in these decays.  By employing the Dalitz plot amplitude analysis 
technique~\cite{dalitz}, the experimental efforts on relevant decay processes combined with the analysis
within the isobar formalism have revealed valuable information on low-energy resonance 
dynamics~\cite{PDG22,HFAG}. Very recently, the Belle~II Collaboration presented a measurement for the decay channels 
 $B^+\to\bar{D}^{(*)0} K^+{K}_S^0$ and $B^0\to D^{(*)-}K^+{K}_S^0$~\cite{2305.01321,2306.07524}.  In addition to 
the four branching fractions for these concerned decays, the $m(K^+ K_S^0)$ distribution of kaon pairs was also provided, 
showing relevant low-mass structures in all four channels~\cite{2305.01321}.

\begin{figure}[tbp]  
\centerline{\epsfxsize=12cm \epsffile{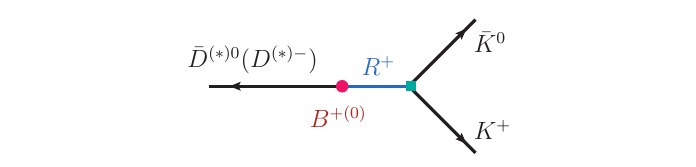}}
\caption{Schematic view of the cascade decays $B^{+}\to \bar{D}^{(*)0} R^+ \to \bar{D}^{(*)0} K^+\bar{K}^0$ and 
              $B^{0}\to D^{(*)-} R^+ \to D^{(*)-} K^+\bar{K}^0$, where $R^+$ stands for the intermediate states 
              $\rho(770,1450)^+$, $a_0(980,1450)^+$ or $a_2(1320)^+$ which decays into $K^+\bar{K}^0$ in this work. }
\label{fig-1}
\end{figure}

Given the presence of an open charm meson $D^{(*)}$ in the final state, these four decay processes measured by 
Belle~II, which have also been previously searched by the Belle experiment~\cite{plb542-171}, are relatively simple and clear 
from a theoretical point of view. One only has to consider the contributions from the tree-level $W$ exchange operators $O_1$ 
and $O_2$ in the effective Hamiltonian $\mathcal{H}_{\rm eff}$~\cite{rmp68.1125} within the framework of the 
factorization method~\cite{zpc34-103}. In the low-mass region, the isospin $I=1$  $K^+ \bar{K}^{0}$ kaon pair emitted in the
$B^+\to\bar{D}^{(*)0} K^+\bar{K}^0$ and $B^0\to D^{(*)-}K^+\bar{K}^0$ decays can be originated from the charged 
intermediate states, $\rho(770)^+$, $a_0(980)^+$, $a_2(1320)^+$ and their excited states, via 
the quasi-two-body mechanism shown schematically in Fig.~\ref{fig-1}. The intermediate state $R^+$ in the figure, which 
decays into the final kaon pair, is generated in the hadronization of the light quark-antiquark 
pair $u\bar{d}$ or can be formed as a dynamically generated state through the meson-meson interactions.

The neutral states $\phi(1020)$, $\omega(782)$ and their excited states will not decay into $K^+ K^{0}_S$ as a result of
charge conservation. The charged $\rho$ resonances are then the expected intermediate states contributing to the $K^+ K^{0}_S$ 
system with spin-parity $J^P=1^-$. In principle, the natural decay mode $\rho(770)\to K\bar{K}$ is blocked because the 
pole mass of the resonance $\rho(770)$ is below the threshold of the kaon pair. However, the virtual 
contribution~\cite{plb25-294,Dalitz62,prd94-072001,plb791-342} from the Breit-Wigner (BW)~\cite{BW-model} 
tail effect of the $\rho(770)$ was found to be indispensable for the productions of kaon pairs in the  processes
$\pi^-p\to K^-K^+n$ and $\pi^+n\to K^-K^+p$~\cite{prd15-3196,prd22-2595}, $\bar p p \to K^+K^-
\pi^0$~\cite{plb468-178,epjc80-453}, $e^+e^- \to K^+K^-$~\cite{pl99b-257,pl107b-297,plb669-217,prd76-072012,
zpc39-13,prd88-032013,prd94-112006,plb779-64,prd99-032001}, and $e^+ e^- \to K^0_{S}K^0_{L}$~\cite{pl99b-261,
prd63-072002,plb551-27,plb760-314,prd89-092002,jetp103-720}. Besides, the mesons $\rho(770)^\pm$ and 
$\rho(1450)^\pm$ are the important intermediate states for the hadronic $\tau$ decays with $K^\pm K^0_S$ in the 
final states~\cite{prd98-032010,prd89-072009,prd53-6037,epjc79-436}. In recent years, the contributions for kaon pairs 
originating from the $\rho$ family of resonances have been explored 
in Refs.~\cite{prd101-111901,prd103-056021,prd103-016002,cpc46-053104} for quasi-two-body $B$ meson decays and in 
Refs.~\cite{prd85-092016,prd93-052018,prd103-114028,prd104.116019} for $D$ meson decays.

The $a_0(980)$ is an experimentally well established scalar state, which has been primarily seen as an 
enhancement in the $\pi\eta$ channel~\cite{prl21.1832}, as well as in the $K\bar{K}$ system near threshold~\cite{plb25-294}.
It has commonly been placed together with the states $f_0(500)$, $K^*_0(700)$ and $f_0(980)$  into a SU$(3)$ 
flavour nonet. The quark-antiquark configuration in the naive quark model for their internal structure cannot explain its true 
nature. In this context, scenarios such as tetraquark states~\cite{prd15.267,prd15.281,plb753-282,plb608-69,npa728-425}, 
molecular states~\cite{prl48.659,epja37-303} and dynamically generated states from meson-meson interactions~\cite{npa620-438,
prl80.3452,prd59.074001,prd60.074023} or a quark-antiquark 
seed~\cite{prd65.114010,plb641-265,prd93.014002,1608.06569} have been 
adopted to describe the mysterious properties of the  $a_0(980)$;  see Refs.~\cite{jpg28-R249,pr389-61,pr397-257,pr454-1,PDG22} 
for reviews in this matter. Conversely, the state $a_0(1450)$, first observed from $\pi\eta$ pair~\cite{plb333-277}, is usually described as a $q\bar{q}$ resonance in the phenomenological studies of Refs.~\cite{prd73.014017,prd75.056001,prd87.114001,epja49-78,prd105.033003}. This resonance $a_0(1450)^+$ is however 
expected to contribute to the kaon pair distribution from the $B^+\to\bar{D}^{(*)0} K^+\bar{K}^0$ and 
$B^0\to D^{(*)-}K^+\bar{K}^0$ decays with a small amount, in view of its tiny decay 
constant~\cite{prd73.014017,plb462-14} and the small ratio between the $K\bar{K}$ decay channel and its dominant
$\omega\pi\pi$ mode~\cite{PDG22,prd78.074023}.

As for the contribution of the isovector tensor meson $a_2(1320)$, we note that it is the ground state of the $a_2$ family 
with quantum numbers $I^GJ^{PC}=1^-2^{++}$ and it can be reasonably understood as a constituent quark-antiquark 
pair within the quark-model~\cite{PDG22}. The transition form factors for the $B$ meson to the $a_2(1320)$ state have 
been obtained in Refs.~\cite{plb695-444,prd83.014008,mpla26-2761,prd100.094005,epjc82-451} within various methods. 
Moreover, the hadronic $B$ meson decays involving a tensor meson $a_2(1320)$ in the final state have been studied 
in Refs.~\cite{epjc22-683,prd67.014002,jpg36-095004,prd83.014007,prd83.034001,prd85.054006,prd86.094001,plb732-36} 
in recent years. The tensor meson $a_2(1700)$, assigned as the first radial excitation of the $a_2(1320)$~\cite{PDG22,prd55.4157} 
state, will not be considered in this work in view of the negligible branching fraction of the decay of the $a_2(1700)$ into 
$K\bar{K}$ pairs~\cite{PDG22,prd90.014001}.

%%%%%%%%%%%%%%%%%%%%%%
%%%%%%%%%%%%%%%%%%%%%%

This paper is organized as follows. In Sec.~\ref{sec-2}, we briefly describe the theoretical framework for obtaining 
the resonance contributions to the decay rates of the $B^{+}\to \bar{D}^{(*)0} R^+ \to \bar{D}^{(*)0} K^+\bar{K}^0$ and 
$B^{0}\to D^{(*)-} R^+ \to D^{(*)-} K^+\bar{K}^0$ processes,  relegating to Appendices~\ref{sec-FMs} and~\ref{sec-appx} the 
specific details of the calculation of the decay amplitudes.    In Sec.~\ref{sec-3}, we present our numerical results of the 
branching fractions for the concerned quasi-two-body decay processes along with some necessary discussions. To test our model, we will also present results for the branching ratios of the $B$ decay processes into a $D^{(*)}$ meson and a pair of pions in the final state. A summary and the conclusions of this work are given in Sec.~\ref{sec-sum}.  

%%<<<<><><><><><><><><><><><><><><><><><><><><><><><><><><><><><><><><>>>>%%
\section{Framework}\label{sec:2}%%%========SEC2 
\label{sec-2}  
%%<<<<><><><><><><><><><><><><><><><><><><><><><><><><><><><><><><><><>>>>%%
In the present paper, we analyze the low-mass enhancement in the distribution of kaon pairs in the final states 
of the $B^+\to\bar{D}^{(*)0} K^+\bar{K}^0$ and $B^0\to D^{(*)-}K^+\bar{K}^0$ decay processes within the 
factorization method. We will specifically concentrate on the resonances 
contributing to the invariant mass region of $m(K^+\bar{K}^0)<1.7$ GeV, adopting the quasi-two-body 
framework for the relevant decays. The quasi-two-body framework based on the perturbative QCD 
(PQCD)~\cite{plb504-6,prd63.054008,prd63-074009,ppnp51-85} approach 
has been discussed in detail in Ref.~\cite{plb763-29} and has been applied to the study of $B$ meson decays in 
Refs.~\cite{prd101-111901,prd103-056021,prd103-016002,cpc46-053104,prd96-093011,npb923-54,
Ma-cpc,jhep2003-162,prd103.013005,epjc80-815,2311.00413} in recent years. 
Parallel analyses for the related three-body $B$ meson decay processes within 
QCD factorization (QCDF) can be found in Refs.~\cite{2007-08881,jhep2006-073,plb622-207,plb669-102,prd79-094005,
prd72-094003,prd88-114014,prd89-074025,prd94-094015,npb899-247,epjc75-536,prd89-094007,prd102.053006}, 
and for other works employing relevant symmetry relations one is referred to Refs.~\cite{plb564-90,prd72-094031,prd72-075013,prd84-056002,plb727-136,plb726-337,prd89-074043,
plb728-579,prd91-014029}.

For the cascade decays $B^+\to\bar{D}^{(*)0} R^+\to\bar{D}^{(*)0} K^+\bar{K}^0$ and $B^0\to D^{(*)-} R^+\to D^{(*)-}K^+\bar{K}^0$,  where the intermediate state $R^+$ stands for $\rho(770,1450)^+$, $a_0(980,1450)^+$ or $a_2(1320)^+$, the 
related effective weak Hamiltonian $\mathcal{H}_{\rm eff}$ accounting for the $\bar{b}\to \bar{c}$ transition is written as~\cite{rmp68.1125}
\begin{equation}  %\mathcal 
 \mathcal H_{\rm eff}= \frac{G_F}{\sqrt2} V^*_{cb} V_{ud} \big[ C_1(\mu) O^c_1(\mu)+C_2(\mu) O^c_2(\mu) \big], \;\;
\label{eff_Ham}
\end{equation}
where $G_F=1.1663788(6)\times10^{-5}$ GeV$^{-2}$~\cite{PDG22} is the Fermi coupling constant, 
$C_{1,2}(\mu)$ are the Wilson coefficients at scale $\mu$, and $V_{cb}$ and $V_{ud}$ are the Cabibbo-Kobayashi-Maskawa~(CKM) matrix elements. The four-quark operators ${O}^c_{1,2}$ are products of two $V-A$ currents, namely 
${O^c_1}=(\bar b d)_{V-A}\,(\bar u c)_{V-A}$ and ${O^c_2}=(\bar b c)_{V-A}\,(\bar u d)_{V-A}$.

With the factorization ansatz, the decay amplitudes for $B^+\to\bar{D}^{(*)0} K^+\bar{K}^0$ and $B^0\to D^{(*)-}K^+\bar{K}^0$ 
are given as~\cite{prd67-034012}
\begin{eqnarray}
  {\mathcal M}(\bar{D}^{(*)0}K^+\bar{K}^0)&=&  \frac{G_F}{\sqrt2} V^*_{cb} V_{ud}
                     \big[  a_2 \langle \bar{D}^{(*)0}|(\bar b c)_{V-A}|  B^+\rangle   %\;\; \nonumber \\ &\times& 
                       \langle K^+\bar{K}^0|(\bar u d)_{V-A}|0\rangle                      \nonumber\\
                      && +  a_1 \langle K^+\bar{K}^0|(\bar b d)_{V-A}|B^+\rangle   %\nonumber \\ &\times& 
                        \langle \bar{D}^{(*)0}|(\bar u c)_{V-A}|0\rangle      \big],  
     \label{eq_das_ch13}\\
{\mathcal M}(D^{(*)-}K^+\bar{K}^0)&=&\frac{G_F}{\sqrt2} V^*_{cb} V_{ud}\, a_2 
                     \langle D^{(*)-}|(\bar b c)_{V-A}| B^0\rangle   %\nonumber\; \\ &\times& 
                         \langle K^+\bar{K}^0|(\bar u d)_{V-A}|0\rangle,
\label{eq-das_ch24}
\end{eqnarray}   
where the effective Wilson coefficients are expressed as $a_1=C_1+C_2/3$ and $a_2=C_2+C_1/3$.

The differential branching fraction ($\mathcal{B}$) for the considered decays is written as~\cite{prd79-094005,appb42.2013,PDG22}
\begin{eqnarray}
   \frac{d\mathcal{B}}{\sqrt{s}\, d\sqrt{s}} = \mathcal{\tau}_B 
                \frac{ \vert\mathbf{p}_{K}\vert \vert\mathbf{p}_{D}\vert}{4(2\pi)^3 m_B^3}  
                \bigg[ \vert\mathcal{M}_S\vert^2        % \nonumber \\ 
               + \frac{1}{3}\vert\mathbf{p}_{K}\vert^2\ \vert\mathbf{p}_{D}\vert^2\vert\mathcal{M}_V\vert^2 \bigg]\!,
\label{eq_dB}
\end{eqnarray}
where the amplitudes $\mathcal{M}_V$ and $\mathcal{M}_S$ are related 
to the vector $\rho(770,1450)^+$ and scalar $a_0(980,1450)^+$ intermediate states, respectively, 
with the help of the Eqs.~(\ref{eq_das_ch13})-(\ref{eq-das_ch24}). 
Here, $\mathcal{\tau}_B~(m_B)$ is the mean lifetime (mass) for the $B$ meson, 
$s=m^2_{K^+\bar{K}^0}$ is the invariant mass square, and
\begin{eqnarray}
     \vert\mathbf{p}_{K}\vert&=&\frac{\sqrt{\left[s-(m_{K^+}+m_{\bar K^0})^2\right]
                 \left[s-(m_{K^+}-m_{\bar K^0})^2\right]}}{2\sqrt s}, \quad
   \label{eq-pK}               
      \\
    \vert\mathbf{p}_{D}\vert&=&\frac{\sqrt{\left[m^2_{B}-(\sqrt{s}+m_{D})^2\right]\left[m^2_{B}
                    -(\sqrt{s}-m_{D})^2\right]}} {2\sqrt s},
 \label{eq-pKpD}
\end{eqnarray}
correspond, respectively, to the magnitude of the momentum of each kaon and that of the bachelor meson $\bar{D}^{(*)0}$ or $D^{(*)-}$, with mass $m_D$, in the rest frame of the intermediate resonance.

%%~~~~~~~~~~~~~~~~~~~~~~Fig.2~~~~~~~~~~~~~~~~~~~~~~~~%%
\begin{figure}[tbp]   %tbp
\centerline{\epsfxsize=14cm \epsffile{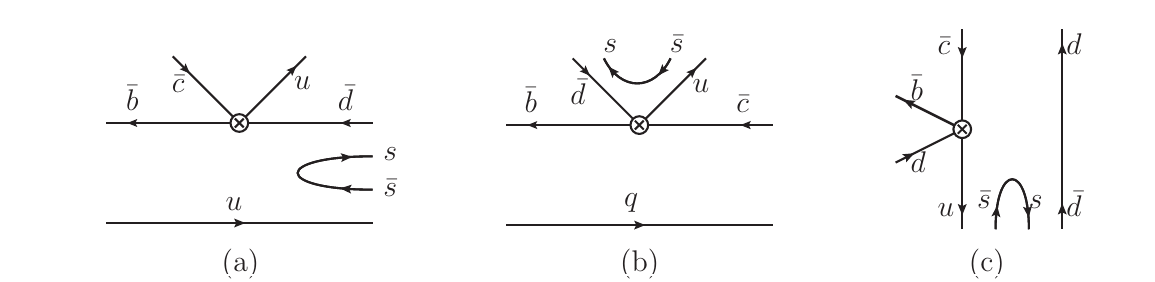}}
\caption{Typical Feynman diagrams for the decays 
              $B^{+}\to \bar{D}^{(*)0} R^+ \to \bar{D}^{(*)0} K^+\bar{K}^0$ and 
              $B^{0}\to D^{(*)-} R^+ \to D^{(*)-} K^+\bar{K}^0$ at quark level, 
              where (a) and (b) are the emission diagrams, (c) is the annihilation one, 
              the quark $q=u$ and $d$ for the $B^+$ and $B^{0}$ processes, 
              respectively, and the symbol $\otimes$ stands for the weak vertex.}
\label{fig2-Feyn}
\end{figure}
%%~~~~~~~~~~~~~~~~~~~~~~Fig.2~~~~~~~~~~~~~~~~~~~~~~~~%%

By combining various contributions from the relevant Feynman diagrams at quark level in Fig.~\ref{fig2-Feyn}, 
the total decay amplitudes for the concerned quasi-two-body decays in the PQCD approach are written as
\begin{eqnarray}
  & &  \mathcal{A}_V(B^+ \to \bar{D}^0 [\rho^+ \to] h h')=\frac{G_F}{\sqrt2}V^*_{cb}V_{ud}    %    \nonumber\\
  %& & \phantom{B^+ \to} \times 
  \big[a_1F_{T\rho} +C_2 M_{T\rho}+a_2F_{TD}+C_1M_{TD} \big]\!,  
  \qquad  \label{DA_Bu}    \\
  & & \mathcal{A}_V(B^0 \to {D^-} [\rho^+ \to] h h')=\frac{G_F}{\sqrt2}V^*_{cb}V_{ud}     %\nonumber\\
  %& & \phantom{B^0 \to} \times 
  \big[a_2F_{TD}+C_1M_{TD} +a_1F_{a\rho}  + C_2 M_{a\rho} \big]\!,  
  \qquad   
\label{DA_B0}
\end{eqnarray}
where $hh'\in\{\pi^+\pi^0, K^+\bar{K}^0\}$. The label $F$ ($M$) denotes that the corresponding decay amplitude comes from 
the factorizable (nonfactorizable) Feynman diagrams, the subscripts $T\rho$ and $TD$ stand for the transition $B\to \rho$ and 
$B\to D$, respectively, and the subscript $a\rho$ is related to the annihilation Feynman diagram of Fig.~\ref{fig2-Feyn}~(c).
The specific expressions in the PQCD approach for these general amplitudes $F$ and $M$ in these decay amplitudes 
are found in Appendix~\ref{sec-FMs}. One should note that the $\mathcal{A}$'s here have a constant factor $(2/m_B)^2$
different from  $\mathcal{M}_V$ in Eq.~(\ref{eq_dB}) because of the different definitions between PQCD and QCDF, see the 
corresponding expression for the differential branching fraction of the former in~\cite{prd103-056021}. 

The quasi-two-body decay amplitudes (\ref{DA_Bu})-(\ref{DA_B0}) are related to the corresponding two-body decay 
amplitude $M_{2B}$ for the $B \to \bar{D}\rho^+$ transition via the relation
\begin{eqnarray}  %^{(\lambda)}
      \mathcal{A}_V= M_{2B}\cdot\frac{\langle hh' \vert \rho^+\rangle}{\mathcal{D}_{\rho^+}(s)},
      \label{rela-2BQ2B}
\end{eqnarray}
where ${\langle hh' \vert \rho^+\rangle}$ stands for the coupling between the $\rho^+$ and the $hh'$ pair. 
Note that the former equation is effectively incorporating the electromagnetic form factor associated to the subprocesses 
$\rho(770,1450)^+ \to \pi^+\pi^0$ and $\rho(770,1450)^+\to K^+\bar{K}^0$ in the corresponding quasi-two-body decays, 
given by~\cite{epjc39-41,prd81-094014,prd86-032013,zpc48-445}
\begin{eqnarray} 
   F_{\pi,K}^{R}(s)=c^{\pi,K}_{R} {\rm BW}_{R}(s)\equiv c^{\pi,K}_{R} \frac{m^2_{R}}{\mathcal{D}_R(s)},
   \label{def-Fpi}
\end{eqnarray}
where the label $R$ represents the resonance, $\rho(770)$ or $\rho(1450)$, and the coefficient 
$c^\pi_{R}=f_R g_{R\pi\pi}/(\sqrt2 m_R)$~\cite{epjc39-41} depends on the corresponding decay constant $f_R$, the coupling 
constant $g_{R\pi\pi}$ and the mass $m_R$. To obtain the coefficient $c^{K}_{R}$ we relay on flavor SU$(3)$ symmetry, 
which establishes $g_{\rho(770)^0 K^+K^-}= g_{\rho(770)^0\pi^+\pi^-}/2$~\cite{epjc39-41}. 
The function $ {\rm BW}_{R}(s)$ stands for a Breit-Wigner shape of the form~\cite{epjc39-41,zpc48-445,prd101.012006}    
\begin{eqnarray}
   {\rm BW}_R\equiv \frac{m^2_{R}}{\mathcal{D}_R(s)}= \frac{m_{R}^2}{m_R^2-s-i m_R \Gamma_{R}(s)}\,,    
   \label{eq-BW}
\end{eqnarray}
with the $s$-dependent width given by
\begin{eqnarray}\label{def-width}
 \Gamma_{R}(s)
             =\Gamma_R\frac{m_R}{\sqrt s} \frac{ \left| \mathbf{p}_{h} \right|^3}{ \left| \mathbf{p}_{h0}\right|^3} 
                X^2(\left|\mathbf{p}_{h} \right| r^R_{\rm BW}),
  \label{eq-sdep-Gamma}
\end{eqnarray}
where $h$  stands for the pion and kaon, respectively, in the $\pi^+\pi^0$ and $K^+\bar{K}^0$ final state pairs.
The magnitude $\left| \mathbf{p}_{h0}\right|$ corresponds to the value of $\left| \mathbf{p}_{h} \right|$ at $s=m^2_R$, 
while $\left| \mathbf{p}_{\pi} \right|$ can be obtained from Eq.~(\ref{eq-pK}) with the replacement of $m_{K^{+,0}}$ 
by $m_{\pi^{+,0}}$. The Blatt-Weisskopf barrier factor~\cite{BW-X} with barrier radius $r^R_{\rm BW}=4.0$ 
GeV$^{-1}$~\cite{prd101.012006} is given by
\begin{eqnarray}
     X(z)=\sqrt{\frac{1+z^2_0}{1+z^2}}\,.
\end{eqnarray}

As for the other two decay amplitudes corresponding to the $B^+$ and $B^0$ decays into final vector mesons $\bar{D}^{*0}$ 
and $D^{*-}$, respectively, they are obtained from Eqs.~(\ref{DA_Bu})-(\ref{DA_B0}) with the replacement of the $D$ meson wave 
function by the $D^*$ one. As has been done in the study of $B$ decays into two vector mesons in the final state, the two-body 
decay amplitudes for $B \to \bar{D}^*\rho^+$ in this work can be decomposed as~\cite{prd76.074018}
%\begin{widetext}
\begin{eqnarray}
M^{(\lambda)}_{2B}
 &=&\epsilon_{\bar{D}^*\mu}^{*}(\lambda)\epsilon_{\rho\nu}^{*}(\lambda) \big[ a
      \,g^{\mu\nu} + {b \over m_{D} \sqrt{s} } P_{B}^\mu P_{B}^\nu         % \nonumber\\
      + i{c \over m_{D}\sqrt{s}} 
      \epsilon^{\mu\nu\alpha\beta} P_{\alpha} P_{3\beta}\big]\!,
  \nonumber \\
 &\equiv&  M_{L}+M_{N}
     \epsilon^{*}_{\bar{D}^*}(\lambda=T)\cdot\epsilon^{*}_{\rho}(\lambda=T)         %\nonumber\\
     + i \frac{M_{T}}{m_{B}^2}
     \epsilon^{\alpha \beta\gamma \eta} \epsilon^{*}_{\rho\alpha}(\lambda)\epsilon^{*}_{D^*\beta}(\lambda) 
     P_{\gamma }P_{3\eta }, \quad
\end{eqnarray}
%\end{widetext}
with three kinds of polarizations of the vector meson, namely, longitudinal ($L$), normal ($N$) and transverse ($T$). 
According to the polarized decay amplitudes, one has the total decay amplitude 
$|\mathcal{A}_V|^2=|A_L|^2+|A_\parallel|^2+|A_\perp|^2$, 
and a longitudinal polarization fraction $\Gamma_L/\Gamma=|A_L|^2/{|\mathcal{A}_V|^2}$, where the amplitudes 
$A_L, A_\parallel$ and $A_\perp$ are related to the two-body amplitudes $M_{L}, M_{N}$ and $M_{T}$, respectively, via Eq.~(\ref{rela-2BQ2B}).
For a detailed discussion, one is referred to Refs.~\cite{prd76.074018,prd45.193,prd46.2969,zpc55-497,prd61.074031}.

%%<<<<><><><><><><><><><><><><><><><><><><><><><><><><><><><><><><><><>>>>%%
\section{Results and Discussions} %%%========SEC3
\label{sec-3}
%%<<<<><><><><><><><><><><><><><><><><><><><><><><><><><><><><><><><><>>>>%%

In this section we present our results for the branching ratios of the decay of $B$ mesons into a charm $D$ or $D^*$ meson and 
a pair of light pseudoscalar mesons. In the numerical calculations, we adopt the decay constants $f_{\rho(770)}=0.216\pm0.003$ 
GeV~\cite{jhep1608-098} and $f_{\rho(1450)}=0.185^{+0.030}_{-0.035}$ GeV~\cite{plb763-29,zpc62-455} for the $\rho(770)$ and 
$\rho(1450)$ resonances, respectively,  and the mean lives $\tau_{B^\pm}=1.638\times 10^{-12}$ s and 
$\tau_{B^0}=1.519\times 10^{-12}$ s for the initial states $B^\pm$ and $B^0$~\cite{PDG22}, respectively. 
The masses for the particles in the relevant decay processes, the decay constants for $B$, $D$ 
and $D^*$ mesons (in units of GeV), and the Wolfenstein parameters for the CKM matrix elements, $A$ and $\lambda$, 
are presented in Table~\ref{tab_con}. We adopt the full widths $\Gamma_{\rho(770)}=149.1\pm0.8$ MeV, 
$\Gamma_{\rho(1450)}=400\pm60$ MeV,  $\Gamma_{a_0(1450)}=265\pm13$ MeV, 
and $\Gamma_{a_2(1320)}=107\pm5$ MeV for the intermediate states involved in this work.

%\vspace{-0.2cm}  %%%%%%%%- Table Cons -%%%
\begin{table}[thb] 
\begin{center}      
\caption{Masses, decay constants (in units of GeV) for relevant states, as well as the Wolfenstein
               parameters for the CKM matrix elements from the {\it Review of Particle Physics}~\cite{PDG22}.  
              The value of $f_{D^*}$ is taken from~\cite{prd96.034524}.}
\setlength{\tabcolsep}{12pt}  %%---%%
\label{tab_con}
\begin{tabular}{l  c c c}\hline\hline
    $m_{B^{\pm}}=5.279$          &$m_{B^{0}}=5.280$          &$m_{D^{*\pm}}=2.010$      &$m_{D^{*0}}=2.007$  \\
    $m_{D^{\pm}}=1.870$         &$m_{D^{0}}=1.865$          &$m_{\pi^{\pm}}=0.140$      &$m_{\pi^{0}}=0.135$  \\ 
    $m_{K^{\pm}}=0.494$         &$m_{K^{0}}=0.498$          &$m_{\rho(770)}=0.775$      &$m_{\rho(1450)}=1.465$ \\
    $m_{a_0(980)}=0.980$       &$m_{a_0(1450)}=1.474$    &$m_{a_2(1320)}=1.318$      &$f_{B}=0.190$  \\   
    $f_{D}=0.212$                      &$f_{D^{*}}=0.2235$            & $A=0.826$                            & $\lambda=0.225$ \\
\hline\hline   
\end{tabular}
\end{center}
\end{table}   
%\vspace{-0.2cm}  %%%%%%%%- Table Cons -%%%

To illustrate the capabilities of the PQCD approach, we first obtain the branching fractions for the quasi-two-body decays  
$B^{+}\to \bar{D}^{(*)0} [\rho(770)^+ \to] \pi^+\pi^0$ and $B^{0}\to D^{(*)-} [\rho(770)^+ \to] \pi^+\pi^0$. Our results,  
displayed in Table~\ref{tab1}, employ the $P$-wave two-pion distribution amplitudes of Ref.~\cite{plb763-29}, the 
$D^{(*)}$ meson wave functions of Refs.~\cite{2311.00413,prd78.014018} and consider $\mathcal{B}_{\rho(770)^+\to\pi^+\pi^0}
\approx 100\%$~\cite{PDG22}. 
%%%%%%%%%%%%%-Table 1-%%
\begin{table}[thb] %thb H
\begin{center}  
\caption{PQCD results for the branching fractions of the quasi-two-body decays 
                $B^{+}\to \bar{D}^{(*)0} [\rho(770)^+ \to] \pi^+\pi^0$ and $B^{0}\to D^{(*)-} [\rho(770)^+ \to] \pi^+\pi^0$.}
\label{tab1}   
\setlength{\tabcolsep}{12pt}  %%---%%
\begin{tabular}{l c c} \hline\hline
  \;Decay modes                                                        &\;Units            
                                                  &PQCD\;                                                                    \\      %& \;Data~\cite{PDG22} \\  
       \hline
  $B^+   \to \bar{D}^0 [\rho(770)^+ \to] \pi^+\pi^0$\;  &$\%$ 
                                             &\;$1.21^{+0.16+0.10+0.05}_{-0.16-0.12-0.06}$     \\      % & $1.34\pm0.18$\; \\ 
%                                                &\;$1.21^{+0.20}_{-0.21}$\;            & $1.34\pm0.18$\; \\ 
  $B^0   \to D^{-}     [\rho(770)^+ \to] \pi^+\pi^0$\;     &$10^{-3}$ 
                                              &\;$7.63^{+0.58+0.97+0.34}_{-0.58-0.73-0.21}$     \\     %  & $7.6\pm1.2$\; \\ 
%                                                 &\;$7.63^{+1.18}_{-0.96}$\;             & $7.6\pm1.2$\; \\ 
  $B^+  \to\bar{D}^{*0}[\rho(770)^+ \to] \pi^+\pi^0$\;  &$10^{-3}$ 
                                              &\;$9.03^{+1.55+0.73+0.51}_{-1.55-0.64-0.46}$      \\     % & $9.8\pm1.7$\; \\ 
 %                                                &\;$9.03^{+1.79}_{-1.74}$\;            & $9.8\pm1.7$\; \\ 
  $B^0   \to D^{*-}    [\rho(770)^+ \to] \pi^+\pi^0$\;     &$10^{-3}$ 
                                             &\;$8.15^{+1.31+0.64+0.03}_{-1.31-0.62-0.07}$       \\     % & $6.8\pm0.9$\; \\ 
%                                                &\;$8.15^{+1.46}_{-1.45}$\;             & $6.8\pm0.9$\; \\ 
\hline\hline
\end{tabular}   
\end{center}
\end{table}
%%%%%%%%%%%%%-Table 1-%%

The calculated branching fractions agree well with the corresponding data 
\begin{eqnarray}
  \mathcal{B}(B^+\to\bar{D}^{0}\rho(770)^+)  &=&   (1.34\pm0.18)\%\,,                 \label{br-2b-Bu2D}       \\
  \mathcal{B}(B^0\to D^{-} \rho(770)^+)         &=&   (7.6\pm1.2)\times10^{-3}\,,    \label{br-2b-Bd2D}      \\
  \mathcal{B}(B^+\to\bar{D}^{*0}\rho(770)^+) &=&   (9.8\pm1.7)\times10^{-3}\,,    \label{br-2b-BuDv}       \\
  \mathcal{B}(B^0\to D^{*-} \rho(770)^+)        &=&   (6.8\pm0.9)\times10^{-3}\,,    \label{br-2b-BdDv}
\end{eqnarray}
in the {\it Review of Particle Physics}~\cite{PDG22}, indicating that the framework employed and the inputs adopted in this 
work are adequate. The branching fraction (\ref{br-2b-Bu2D}) for $B^+\to\bar{D}^{0}\rho(770)^+$ was averaged 
in~\cite{PDG22} from the data $ (1.35\pm0.12\pm0.15)\%$ and  $(1.3\pm0.4\pm0.4)\%$ presented by CLEO and ARGUS in
Ref.~\cite{CLEO:1994mwq} and Ref.~\cite{ARGUS:1990jet}, respectively. Very recently, the Belle II Collaboration measured 
the decay $B^-\to D^0\rho(770)^-$ using data collected with the Belle II detector, its branching fraction was determined to 
be $(0.939\pm0.021({\rm stat})\pm 0.050({\rm syst}))$\% by restricting the $\pi^-\pi^0$ invariant mass to a $300$ MeV range 
centered at the $\rho(770)^-$ mass pole~\cite{2404.10874}.  This measurement is smaller than the value in Eq.~(\ref{br-2b-Bu2D}), 
but it is still in agreement with our result in Table~\ref{tab1} within the uncertainties.

With the help of the kaon form factor $F_{K^+\bar{K}^0}(s)$ discussed in detail in~\cite{prd103-056021}, we obtain the 
concerned branching fractions of the $B$ mesons into a $D$ or $D^*$ meson and a pair of kaons for the quasi-two-body 
processes $\rho(770)^++\rho(1450)^+ \to K^+\bar{K}^0$. Our results are displayed in Table~\ref{tab2}.
%%%%%%%%%%%%%-Table 2-%% PQCD Q2B with KK %%%%
\begin{table}[H] %thb H
\begin{center}                                                                   
\caption{PQCD predictions for the branching fractions of the concerned quasi-two-body decays with 
              the subprocess $ \rho^+ \to K^+\bar{K}^0$, here $\rho^+=\rho(770)^++\rho(1450)^+$.}  
\label{tab2}   
\setlength{\tabcolsep}{12pt}  
\begin{tabular}{l c c } \hline\hline
     \quad Mode                                                                          &    \;\;Unit\;\;          & \;\;${\mathcal B}$                    \\         \hline             
     $B^+ \to \bar{D}^0 [\rho^+\to]K^+\bar{K}^0$\;         &$\; 10^{-4} \;$         
                                                                                                     &\;$1.68^{+0.20+0.17+0.12}_{-0.20-0.15-0.12}$         \\ 
     $B^0 \to        {D}^- [\rho^+\to]K^+\bar{K}^0$\;            &$\; 10^{-4} \;$         
                                                                                                     &\;$0.98^{+0.06+0.13+0.06}_{-0.06-0.12-0.06}$         \\ 
     $B^+ \to \bar{D}^{*0}[\rho^+\to]K^+\bar{K}^0$\;      &$\; 10^{-4} \;$         
                                                                                                     &\;$1.33^{+0.21+0.11+0.05}_{-0.21-0.11-0.07}$          \\      
     $B^0 \to        {D}^{*-}[\rho^+\to]K^+\bar{K}^0$\;        &$\; 10^{-4} \;$        
                                                                                                     &\;$1.16^{+0.19+0.08+0.02}_{-0.19-0.09-0.02}$          \\      
\hline\hline
\end{tabular}   
\end{center}
%\vspace{-0.5cm} 
\end{table}
%%%%%%%%%%%%%-Table 2-%% PQCD Q2B with KK %%%%

In the results for the branching fractions shown in Tables~\ref{tab1}-\ref{tab2}, the first source of the error corresponds to 
the uncertainties of the shape parameter $\omega_B=0.40\pm0.04$ of the $B^{\pm,0}$ wave functions, while the 
the Gegenbauer moments $C_D=0.6\pm0.15$ or  $C_{D^*}=0.5\pm0.10$ present in the $D$ or $D^*$ wave functions~\cite{2311.00413}  contribute to the second source of error. The third one is induced by the Gegenbauer moments 
$a_R^{0}=0.25\pm0.10$, $a_R^t=-0.60\pm0.20$ and $a_R^s=0.75\pm0.25$~\cite{plb763-29} present in the wave functions of 
the intermediate states. The other errors for the PQCD predictions in this work, which come from the uncertainties of the 
masses and the decay constants of the initial and final states, from the uncertainties of the Wolfenstein parameters, etc., 
are small and have been neglected.

Comparing our calculated branching rations of Table~\ref{tab2} with the measured results (in units of $10^{-4}$)~\cite{2305.01321}
\begin{eqnarray}
    \mathcal{B}(B^+\to\bar{D}^0 K^+K_S^0)   &=& 1.89\pm 0.16\pm 0.10\,,            \label{resu_ch1}      \\ 
    \mathcal{B}(B^0\to D^-K^+K_S^0)            &=& 0.85\pm 0.11\pm 0.05\,,            \label{resu_ch2}      \\
    \mathcal{B}(B^+\to \bar{D}^{*0}K^+K_S^0)&=& 1.57\pm 0.27\pm 0.12\,,            \label{resu_ch3}     \\
    \mathcal{B}(B^0\to D^{*-}K^+K_S^0)        &=& 0.96\pm 0.18\pm 0.06\,, \quad  \label{resu_ch4} 
\end{eqnarray}
and taking into account that half of the $\bar{K}^0$ or $K^0$ goes to $K_S^0$, we conclude that an important fraction of 
the decays $B^+\to\bar{D}^{(*)0} K^+\bar{K}^0$ and $B^0\to D^{(*)-}K^+\bar{K}^0$ proceeds through the
intermediate states $\rho(770)^+$ and $\rho(1450)^+$, but there is still room for other contributions. 

One could argue that the resonance $\rho(770)^+$, as a virtual bound state~\cite{Dalitz62,plb25-294}, will not completely 
exhibit its properties in a quasi-two-body cascade decay like  $B^0 \to {D}^- [\rho(770)^+\to]K^+\bar{K}^0$, since the 
invariant masses of the emitted kaon pairs exclude the region around the $\rho(770)$ pole mass. However, as we will show 
below, the width of this resonance renders its contribution quite sizable in the energy region of interest. It is therefore 
important to consider explicitly the subthreshold resonances in the analysis of the branching ratios, even if they contribute 
via the tail of their mass distribution. In other words, experimental analyses or theoretical studies of  three-body $B$ meson 
decay process should not attribute as nonresonant $K\bar K$ invariant mass strength the specific known contribution from 
a certain resonant state like the $\rho(770)$.

%%~~~~~~~~~~~~~~~~~~~~~~Fig.3~~~~~~~~~~~~~~~~~~~~~~~~%%
\begin{figure}[tbp]  
\centerline{\epsfxsize=8cm \epsffile{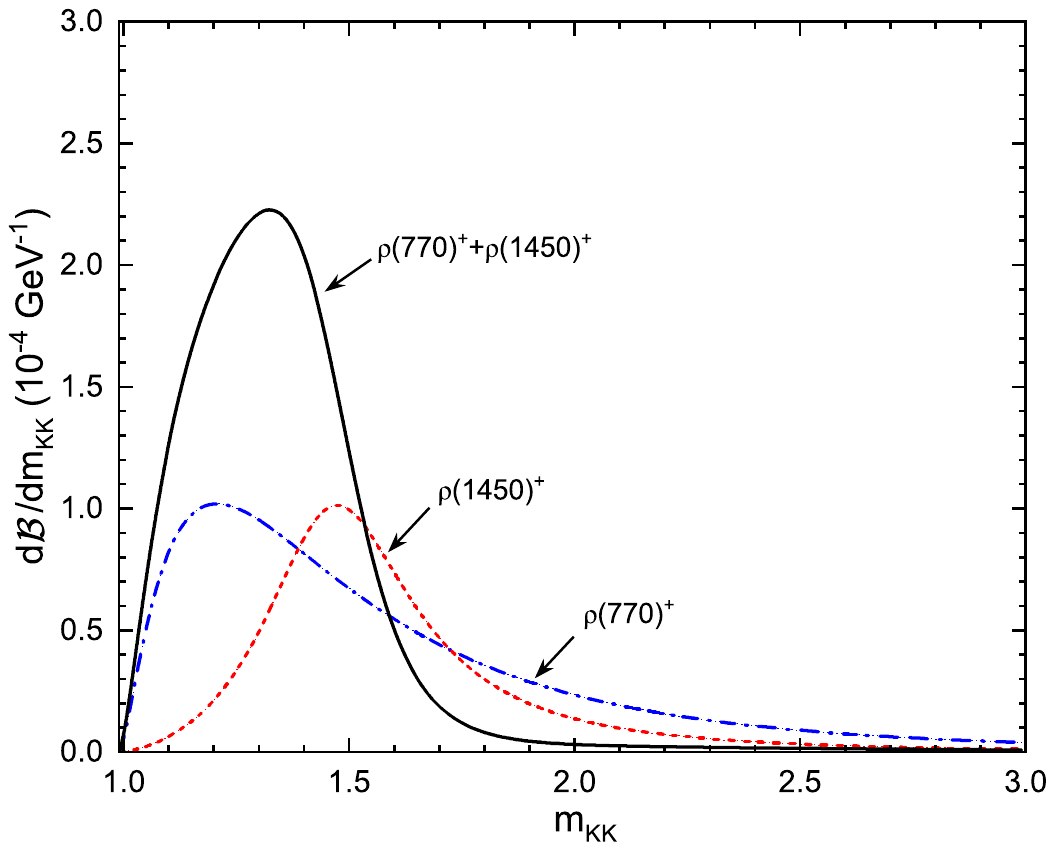}}
\caption{The differential branching fraction for the quasi-two-body decay $B^0 \to {D}^-\rho^+\to {D}^-K^+\bar{K}^0$,
              with the intermediate $\rho^+\in \{\rho(770)^+, \rho(1450)^+, \rho(770)^++\rho(1450)^+\}$.}
\label{fig_dbf_ch2}
\end{figure}
%%~~~~~~~~~~~~~~~~~~~~~~Fig.3~~~~~~~~~~~~~~~~~~~~~~~~%%

%%~~~~~~~~~~~~~~~~~~~~~~Fig.4~~~~~~~~~~~~~~~~~~~~~~~~%%
\begin{figure}[tbp]  
\centerline{\epsfxsize=8cm \epsffile{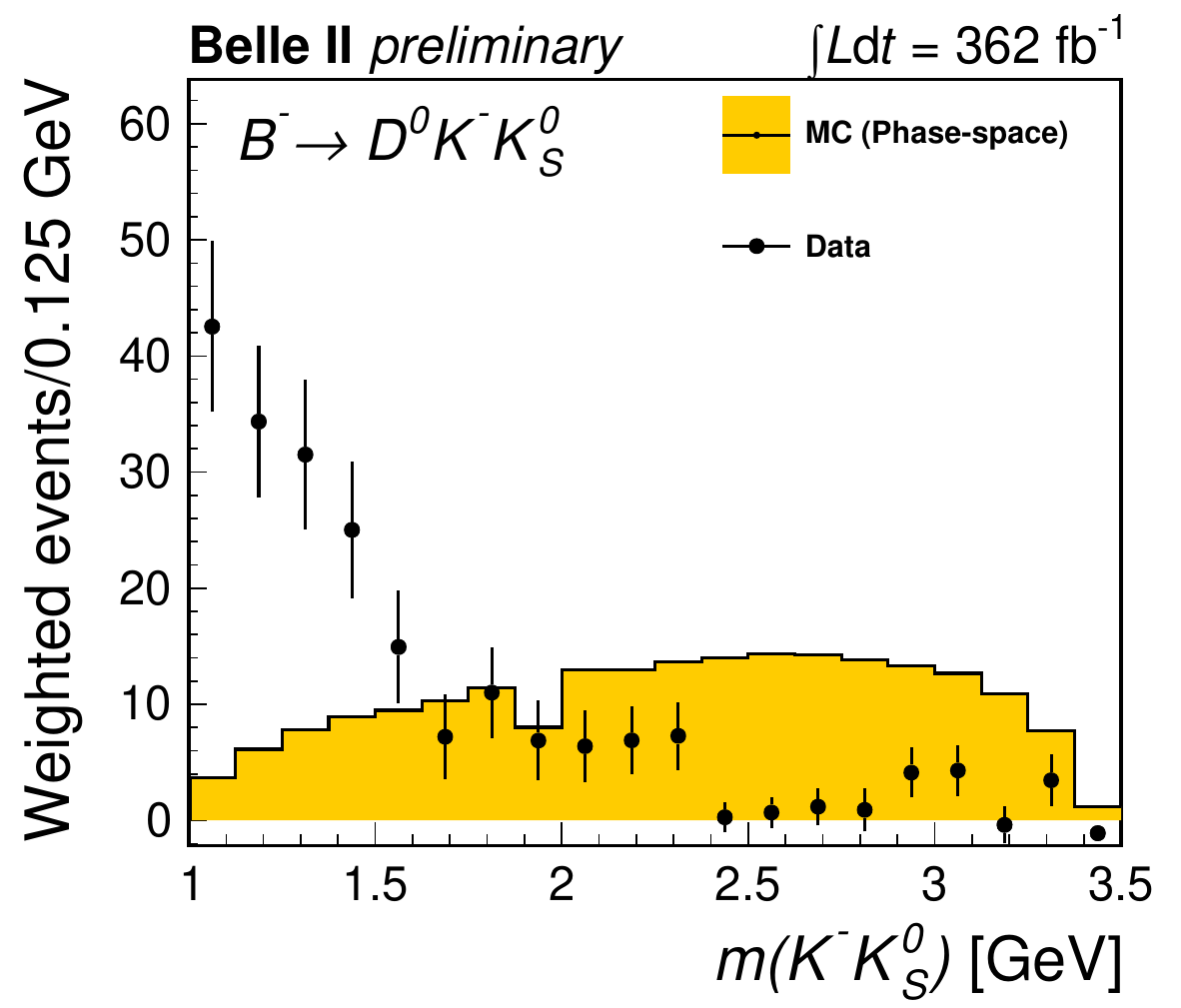}}
\vspace{0.2cm}
\centerline{\epsfxsize=8cm \epsffile{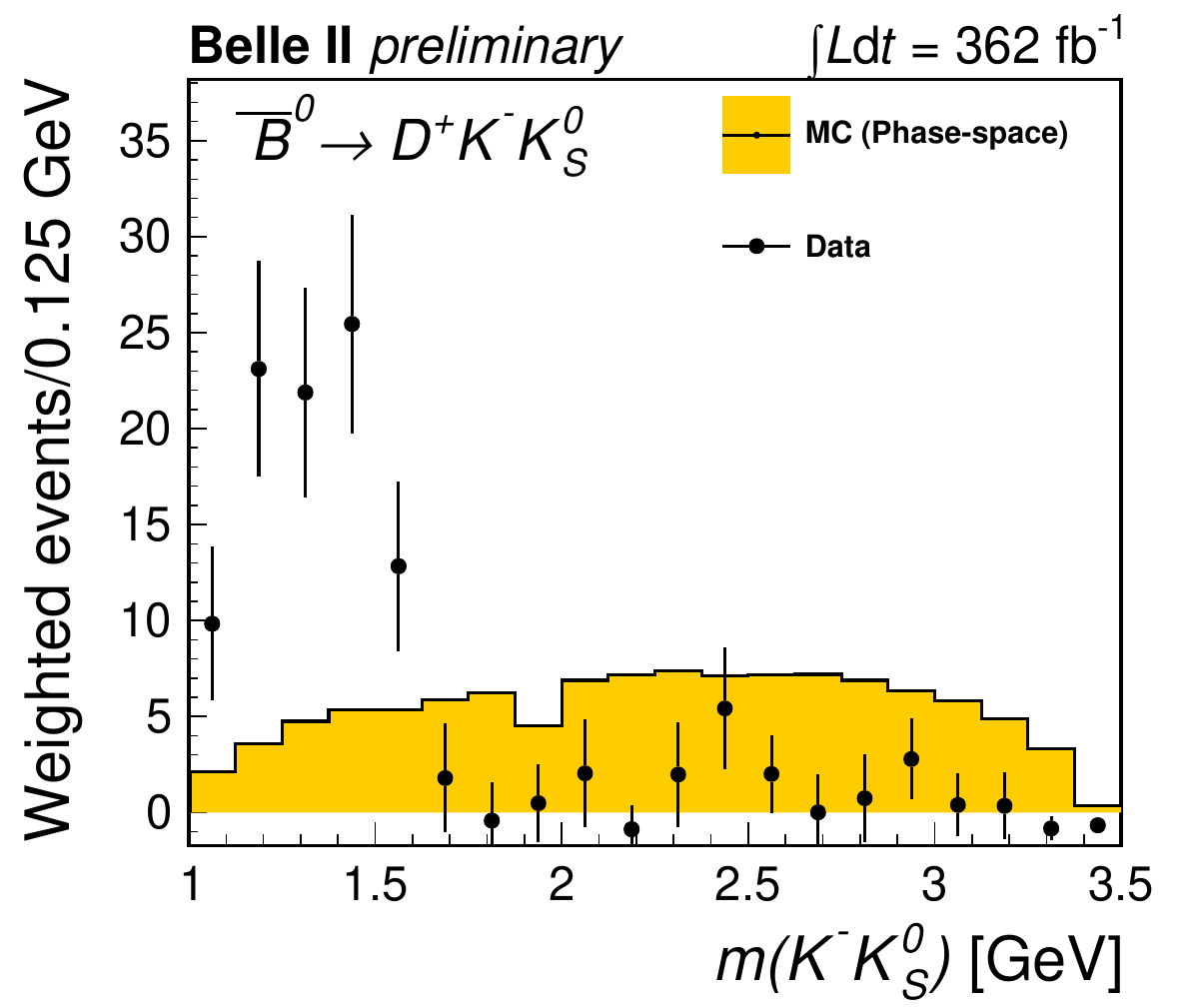}}
\caption{The distribution of $m(K^-{K}^0_S)$ for $B^- \to {D}^0 K^-{K}^0_S$ (top) 
              and $\bar{B}^0 \to {D}^+K^-{K}^0_S$ (bottom) decays from Belle II~\cite{2305.01321}.
             }
\label{figs_Belle}
\end{figure}
%%~~~~~~~~~~~~~~~~~~~~~~Fig.4~~~~~~~~~~~~~~~~~~~~~~~~%%

To make this point more evident, we show in Fig.~\ref{fig_dbf_ch2} the differential branching fraction for the quasi-two-body 
decay $B^0 \to {D}^-\rho^+\to {D}^-K^+\bar{K}^0$. The dashed line with a peak at about $1.465$ GeV reveals the 
contribution from the $\rho(1450)^+$, while the dash-dot line, depicting the contribution of the $\rho(770)^+$, presents 
a bump around $1.2$ GeV, which shall not be claimed experimentally as a resonant state with quite a large decay width. 
This bump is actually formed by the BW tail of the $\rho(770)^+$ along with the phase space factor of Eq.~(\ref{eq_dB}).  
The interference between the BW expressions for $\rho(770)^+$ and $\rho(1450)^+$ is constructive in the region before 
the pole mass of the  $\rho(1450)^+$ and destructive after it as a result of the sign difference between 
$c^{K}_{\rho(770)}=1.247\pm0.019$ and $c^{K}_{\rho(1450)}=-0.156\pm0.015$~\cite{prd103-056021} in Eq.~(\ref{def-Fpi}). 
Note that the theoretical distribution has the same pattern in the low-mass region of the kaon pair as  that shown in the 
bottom panel of Fig.~\ref{figs_Belle} for the three-body decay $\bar{B}^0 \to {D}^+K^-{K}^0_S$. This comparison reflects 
the dominant contributions for this decay coming from the intermediate states $\rho(770)^+$ and $\rho(1450)^+$.

The higher mass resonance $\rho(1700)^+$ as an intermediate state could also decay into 
$K^+\bar{K}^0$~\cite{prd101-111901,prd103-056021,cpc46-053104} and hence contribute to the 
$B^+\to\bar{D}^{(*)0} K^+\bar{K}^0$ and $B^0\to D^{(*)-}K^+\bar{K}^0$ decays.  
Take the case $B^0 \to {D}^-\rho(1700)^+\to {D}^- K^+\bar{K}^0$ as an example. With the coefficient 
$c^{K}_{\rho(1700)}=-0.083\pm0.019$~\cite{prd103-056021}, its branching fraction was predicted to be 
$\mathcal{B}=4.00^{+3.36}_{-2.26}\times10^{-5} $ in~\cite{cpc46-053104}, which represents about a
$20\%$ of the total branching fraction for $B^0\to D^-K^+\bar{K}^0$ when comparing with the result of 
Eq.~(\ref{resu_ch2}) and neglecting the large error from the PQCD prediction.  However,  the $m(K^-{K}^0_S)$ 
distributions from the $B^- \to {D}^0 K^-{K}^0_S$ and $\bar{B}^0 \to {D}^+K^-{K}^0_S$ decays
measured by Belle  II~\cite{2305.01321} , see Fig.~\ref{figs_Belle}, show no prominent enhancement around the mass of 
the $\rho(1700)^+$.  In view of the fact that $\mathcal{B}(a_2(1700)\to K\bar{K})=(1.9\pm1.2)\%$~\cite{PDG22} in the 
same region, the explanation for the lack of structure possibly lies in:   i) the interference between the $\rho(1700)^+$ and 
the $\rho(770)^+ + \rho(1450)^+$ contributions being destructive around $1.7$ GeV; ii) the coefficient $c^{K}_{\rho(1700)}$ 
in~\cite{prd103-056021} being highly overrated, since one can also find sensibly smaller values in the literature, namely 
$-0.015\pm0.022$ in~\cite{epjc39-41} and $-0.028\pm0.012$ in~\cite{prd81-094014}.

Near the threshold of the kaon pair, one finds remarkable enhancements in the $m(K^-{K}^0_S)$ distributions 
for the decays $B^+\to\bar{D}^{0} K^+{K}^0_S$ and $B^+\to\bar{D}^{*0} K^+{K}^0_S$ from Belle II~\cite{2305.01321}, 
but not for $B^0\to D^{-}K^+\bar{K}^0$ or $B^0\to D^{*-}K^+\bar{K}^0$. The invariant kaon pair mass around $1$ GeV
is the energy region of the state $a_0(980)$, but we do not expect the same strength of the $a_0(980)$ contributions
in the $B^+\to\bar{D}^{(*)0} K^+\bar{K}^0$ and the $B^0\to D^{(*)-}K^+\bar{K}^0$ processes, since their decay mechanisms
proceed through different quark-type Feynman diagrams, shown in Fig.~\ref{fig2-Feyn}, as explained in the following.
The annihilation Feynman diagrams represented by Fig.~\ref{fig2-Feyn} (c) will only contribute to the decays 
$B^{0}\to D^{(*)-}R^+\to D^{(*)-}K^+\bar{K}^0$, and the contributions are highly suppressed when comparing 
with the those from the emission diagrams of Figs.~\ref{fig2-Feyn} (a) and  (b).                                                                                                                                                                                                                                                                                                                                                                                                                                                                                                                                                                                                                      

For the decays $B^0\to D^{(*)-}K^+\bar{K}^0$ with subprocess $a_0(980)^+\to K^+\bar{K}^0$, the contribution of the 
Feynman diagrams of Fig.~\ref{fig2-Feyn} (b)  is small. 
This is because the matrix element $\langle K^+\bar{K}^0|(\bar u d)_{V-A}|0\rangle$ with 
$a_0(980)^+$ as the intermediate state depends on the tiny decay constant $f_{a_0(980)}\approx1.1$MeV~\cite{plb462-14,
prd63.074017,prd105.033006,jhep0106-067}. This qualitatively explains why we don't see a remarkable enhancement of 
events around $1$ GeV in the $m(K^-{K}^0_S)$ distribution from the decay $B^0\to D^{(*)-}K^+\bar{K}^0$~\cite{2305.01321}, 
depicted in the bottom panel of Fig.~\ref{figs_Belle}.
But the amplitudes for the decay $B^+\to\bar{D}^{(*)0} K^+\bar{K}^0$ receive contribution not only from 
Fig.~\ref{fig2-Feyn}~(b), with an $a_0(980)^+$ being emitted, but also from the diagram of Fig.~\ref{fig2-Feyn}~(a) which is a 
$B^+\to a_0(980)^+$ transition with an emitted $\bar{D}^{(*)0}$. Despite being a colour suppressed Feynman 
diagram~\cite{plb519-50}, the hierarchical relation $f_{\bar{D}^{(*)0} }\gg f_{a_0(980)}$ makes the 
contribution from Fig.~\ref{fig2-Feyn}~(a) much larger than that from Fig.~\ref{fig2-Feyn}~(b) for the decays 
$B^+\to\bar{D}^{(*)0} K^+\bar{K}^0$ with the subprocess $a_0(980)^+\to K^+\bar{K}^0$.

Let us now proceed to the explicit numerical calculation. 
Within the naive factorization approach, the evaluation of the decay amplitude for the $B^+\to\bar{D}^{0} K^+\bar{K}^0$ decay with the subprocesses 
$a_0(980,1450)^+\to K^+\bar{K}^0$ can be found in Appendix~\ref{sec-appx}. With Eq.~(\ref{da_Bua0}) and the inputs from 
the {\it Review of Particle Physics}~\cite{PDG22}, we obtain a branching fraction $\mathcal{B}=1.56\times10^{-5}$ for 
the quasi-two-body decay  $B^+\to\bar{D}^{0}a_0(980)^+ \to\bar{D}^{0} K^+\bar{K}^0$, which corresponds to 
a value $\mathcal{B}(B^+\to\bar{D}^{0}a_0(980)^+)=1.07\times 10^{-4}$ for the two-body decay. Likewise, we obtain 
$\mathcal{B}=0.72\times10^{-5}$ for $B^+\to\bar{D}^{0}a_0(1450)^+ \to\bar{D}^{0} K^+\bar{K}^0$, 
where we have employed $F_0^{B\to a_0(1450)}(0)=0.26$~\cite{prd69.074025} and 
$\Gamma(a_0(1450)\to K\bar{K})/\Gamma(a_0(1450)\to\omega\pi\pi)\approx0.082$~\cite{PDG22}. In order to check the 
reliability of the method we adopted here, the measured channel $B^0\to D^+_s a_0(980)^-$ is studied as a reference. This is a process with a $B^0\to a_0(980)^-$ transition and an emitted $D^+_s$ state.  Within naive factorization, we find
$\mathcal{B}(B^0\to D^+_s a_0(980)^-)=1.93\times 10^{-5}$. This branching fraction is very
close to the upper limit $1.9\times 10^{-5}$ at $90$\% C.L. presented by the $BABAR$ Collaboration in Ref.~\cite{prd73.071103} 
assuming $\mathcal{B}(a_0(980)^+\to\eta\pi^+)$ to be $100$\%, but it is much smaller than the prediction 
$\mathcal{B}=4.81^{+2.19}_{-1.79}\times 10^{-5}$ in~\cite{prd95.016011} within PQCD for the decay 
$B^0\to D^+_s a_0(980)^-$. However, taking into account 
$\Gamma(a_0(980)\to K\bar{K})/\Gamma(a_0(980)\to \pi\eta)=0.172\pm0.019$~\cite{PDG22}, one has  
$\mathcal{B}(a_0(980)^+\to\eta\pi^+)\approx 0.85$ and this will change the upper limit in~\cite{prd73.071103}
for $B^0\to D^+_s a_0(980)^-$ up to $2.24\times 10^{-5}$ at $90$\% C.L., which is still much smaller than the prediction 
in~\cite{prd95.016011}, hinting that the PQCD approach is possibly not appropriate for the study of the $B^0\to D^+_s a_0(980)^-$ decay
with the $B \to a_0(980)$ transition.

When we put the contributions from $a_0(980,1450)^+\to K^+\bar{K}^0$ and $\rho(770,1450)^+\to K^+\bar{K}^0$ 
for the decay $B^+\to\bar{D}^{0} K^+\bar{K}^0$ together, the resulting differential branching fraction does not have the
shape shown in the top panel of Fig.~\ref{figs_Belle}. 
The contribution from the scalar intermediate state $a_0(980)^+$ is far from what would be required to overcome the 
peak of the $\rho(770,1450)^+$ distribution in order to reproduce the enhancement near the threshold of 
$K^+\bar{K}^0$ pairs measured experimentally. The shape of the measured $B^+\to\bar{D}^{0} K^+\bar{K}^0$ 
differential branching fraction would only be obtained with a branching fraction $\mathcal{B}\approx 4.5\times 10^{-4}$ 
for the quasi-two-body decay  $B^+\to\bar{D}^{0}a_0(980)^+ \to\bar{D}^{0} K^+\bar{K}^0$, which is beyond the total 
branching fraction for $B^+\to\bar{D}^{0} K^+\bar{K}^0$ decay. This situation is probably indicating the existence of large nonresonant contributions to the $B^+\to\bar{D}^{0} K^+\bar{K}^0$ decay around the threshold of the kaon pair or other unknown sources. Note that the interference between $\rho(770)^+$ and $\rho(1450)^+$ could reduce the corresponding branching fractions in Table~\ref{tab2} through an appropriate complex phase difference between their respective BW expressions. 
This would alleviate the requirement of an enhanced contribution from the $a_0(980)^+$.
For example, a factor of $e^{i\pi/4}$ before the BW of the $\rho(1450)^+$ will produce half of the $B^+ \to \bar{D}^0 [\rho^+\to]K^+\bar{K}^0$ branching fraction listed in Table~\ref{tab2}. But such an universal 
phase difference will also make the branching fractions of the decays  $B^0 \to  {D}^{(*)-}[\rho^+\to]K^+\bar{K}^0$  
decrease by half in Table~\ref{tab2}, which is not desirable.

Let us mention that, in the amplitude analysis of the decay $D_{s}^{+} \rightarrow \pi^{+}\pi^{0}\eta$, an unexpected large branching fraction 
$(1.46\pm0.15_{\rm stat.}\pm0.23_{\rm sys.})\%$ was measured for 
$D_{s}^{+} \rightarrow a_{0}(980)^{+(0)}\pi^{0(+)}, a_{0}(980)^{+(0)} \to \pi^{+(0)}\eta$ by the BESIII 
Collaboration~\cite{prl123.112001}, which was successfully interpreted via the rescattering processes 
$K\bar{K}\to a_{0}(980)\to \pi\eta$ or $\pi\eta^{(\prime)}\to a_{0}(980)\to \pi\eta$ with the triangle diagrams 
suppression in Refs.~\cite{plb803-135279,epjc80-895,epjc80-1041,2102-05349}. But one should note that the above decay process is 
quite different when compared with the $B^+\to\bar{D}^{(*)0} K^+\bar{K}^0$ and $B^0\to D^{(*)-}K^+\bar{K}^0$ decays studied here.
For the three-body decay $D_{s}^{+} \rightarrow \pi^{+}\pi^{0}\eta$, the intermediate state $a_{0}(980)$ can 
only be generated by the final state interactions; the $c\bar{s}$ quark pair in the initial state $D_{s}^{+}$ can 
not produce $D_{s}^{+}\to  a_{0}(980)$ transitions directly.

We finally discuss the contribution of the tensor resonance $a_2(1320)$. We note that this resonance
decays into a kaon pair with a small branching fraction of about 5\%~\cite{PDG22}. 
Taking into account that the tensor states cannot be generated from a $V-A$ current~\cite{prd105.093006}, 
we do not expect to have considerable contributions from the subprocess $a_2(1320)^+\to K^+\bar{K}^0$ 
for the decays $B^0\to D^{(*)-}K^+\bar{K}^0$ by the related decay amplitudes in Eq.~(\ref{eq-das_ch24}). 
In Ref.~\cite{prd90.072003}, the quasi-two-body decay $B^0_s \to \bar{D}^0 [\bar{K}^*_2(1430)^0\to]K^-\pi^+$ 
was measured with the branching fraction $\mathcal{B}=(3.7\pm1.4)\times 10^{-5}$, which means 
$\mathcal{B}=(1.1\pm0.4)\times 10^{-4}$~\cite{PDG22} for the corresponding two-body process 
$B^0_s \to \bar{D}^0 \bar{K}^*_2(1430)^0$. %B=1.124*10^{-4}% from theory.
With this measured branching fraction and the replacement of a $s$-quark by a $u$-quark, it is easy to predict 
the branching fraction $\mathcal{B}=(0.99\pm0.37)\times10^{-4}$ for the decay $B^+\to\bar{D}^0 a_2(1320)^+$ 
within SU($3$) flavor symmetry and employing the form factors $A_0$ in~\cite{plb695-444} 
for the $B\to a_2$ and $B_s\to K^*_2$ transitions. This predicted value is consistent with 
the corresponding theoretical results in~\cite{prd67.014011,prd86.094001,prd83.014007}. However, when taking into account the 
branching fraction $\mathcal{B}(a_2(1320)^+\to K^+\bar{K}^0)=4.9\pm0.8$\%~\cite{PDG22}, the contribution from 
$a_2(1320)^+$ to the $B^+\to\bar{D}^{0} K^+\bar{K}^0$  process is negligible. The decay $B^+\to\bar{D}^{*0} a_2(1320)^+$ 
shares the same Feynman diagrams with $B^+\to\bar{D}^{0} a_2(1320)^+$ at quark level, and it is reasonable to infer a 
insignificant branching fraction for the decay $B^+\to\bar{D}^{*0} K^+\bar{K}^0$ with the subprocess 
$a_2(1320)^+\to K^+\bar{K}^0$ as well.

%%<<<<><><><><><><><><><><><><><><><><><><><><><><><><><><><><><><><><>>>>%%
\section{Summary}%%%========SEC-Sum
\label{sec-sum}
%%<<<<><><><><><><><><><><><><><><><><><><><><><><><><><><><><><><><><>>>>%%
To sum up, the Belle~II Collaboration presented a measurement for the $B^+\to\bar{D}^{(*)0} K^+\bar{K}^0$ and 
$B^0\to D^{(*)-}K^+\bar{K}^0$ decays very recently, where the bulk of the observed $m(K^+ K_S^0)$ distribution 
was located in the low-mass region of the kaon pair, showing structures in all four decay channels. In this work we have 
presented a theoretical calculation of these decays within the factorization method. We have focused on exploring the 
region of kaon pair invariant masses  $m(K^+\bar{K}^0)<1.7$ GeV. The resonance contributions from vector intermediate 
states $\rho(770,1450)^+$ have been found to dominate the branching fractions for the three-body decays 
$B^+\to\bar{D}^{(*)0} K^+\bar{K}^0$ and $B^0\to D^{(*)-}K^+\bar{K}^0$, representing roughly half of the total 
branching fractions of the corresponding decay channels.  The role of the tensor $a_2(1320)^+$ was analyzed and 
found to give negligible contributions to the branching fractions of these four decay processes and the contribution 
of the state $a_0(980)^+$ turned out to be less important than expected in the $m(K^+ K_S^0)$ region near the threshold 
of the kaon pair. As a result of our study, we conclude that the enhancement of events in the kaon pair distribution near 
threshold observed in the $B^+\to\bar{D}^{0} K^+\bar{K}^0$ and $B^+\to\bar{D}^{*0} K^+\bar{K}^0$ decay processes 
can not be interpreted as the resonance contributions from the $a_0(980)^+$ meson. The nonresonant contributions 
are probably governing the formation of the kaon pair in $B^+\to\bar{D}^{(*)0} K^+\bar{K}^0$ near the threshold of 
$K^+\bar{K}^0$, and hence deserve  further examination both from the theoretical and the experimental sides.

%-----------------------%
\begin{acknowledgments}
We thank V. Bertacchi, K. Trabelsi and A. Feijoo for valuable discussions. 
This work was supported in part by the National Natural Science Foundation of China under Grants 
No. 12205148 and No. 11947011,  the Fund for Shanxi ``1331 Project" Key Subjects Construction, 
the Natural Science Foundation of Jiangsu Province under Grant No. BK20191010, 
and the Qing Lan Project of Jiangsu Province. This research has also been supported from the projects CEX2019-000918-M (Unidad de Excelencia ``Mar\'{\i}a de Maeztu") and PID2020-118758GB-I00, financed by the Spanish MCIN/ AEI/10.13039/501100011033/, as well as by the EU STRONG-2020 project, under the program H2020-INFRAIA-2018-1 grant agreement no. 824093.
\end{acknowledgments}

%%<<<<><><><><><><><><><><><><><><><><><><><><><><><><><><><><><><><><>>>>%%
\appendix%%%========SEC-Appx

\section{General amplitudes for $B \to \bar{D}^{(*)} \rho\to\bar{D}^{(*)} K^+\bar{K}^0$ decays}  
\label{sec-FMs}
The wave functions for $B$, $D$ and $D^*$ mesons and the corresponding inputs are the same as they in 
Ref.~\cite{2311.00413}. The kaon and pion timelike form factors are referred to the Section II of Ref.~\cite{prd103-056021}.
With the subprocesses $\rho^+\to K^+\bar{K}^0$, where $\rho$ is $\rho(770)$ or $\rho(1450)$, the specific expressions 
in PQCD approach for the Lorentz invariant decay amplitudes of these general amplitudes $F$ and $M$ for 
$B \to \bar{D}^{(*)} \rho\to\bar{D}^{(*)} K^+\bar{K}^0$ decays are given as follows:

The amplitudes from Fig.~\ref{fig2-Feyn} (a) for the decays with a pseudoscalar $\bar{D}^0$ or $D^-$ meson 
in the final states are given as
%\begin{widetext}
%%----------------------------------------------------------
\begin{eqnarray}
F_{T\rho}&=&8\pi C_F m^4_B f_{D}\int dx_B dx\int b_B db_B b db \phi_B \big\{\big[[r^2-\bar{\zeta}(x(r^2-1)^2+1)]\phi^0-\sqrt{\zeta}[(r^2+\bar{\zeta}\nonumber\\
&& +2\bar{\zeta}x(r^2-1))\phi^s -(r^2-1)\bar{\zeta} (2 x(r^2-1) +1)-r^2)\phi^t\big] E_e(t_a)h_a(x_B,x,b,b_B)S_t(x) \nonumber\\
&&+\big[(r^2-1)[\zeta \bar{\zeta}-r^2 (\zeta-x_B)]\phi^0-2 \sqrt{\zeta}[\bar{\zeta}-r^2 (x_B-2 \zeta+1)]\phi^s \big]\nonumber\\
&&\times E_e(t_b)h_b(x_B,x,b_B,b)S_t(|x_B-\zeta|)\big\},
\label{eq:f01}
\end{eqnarray}
%%--------------
\begin{eqnarray}
M_{T\rho}&=&16\sqrt{\frac{2}{3}} \pi C_F m_B^4 \int dx_B dx dx_3 \int b_B db_B b_3 db_3 \phi_B \phi_D\big\{\big[
-[(\bar{\zeta}+r^2) ((r^2-1) (x_3 \bar{\zeta}+x_B) \nonumber\\
&&+r^2 (\zeta x-1)-\zeta (x+1)+1)+r r_c(r^2-\bar{\zeta})]\phi^0-\sqrt{\zeta}[(r^2 (\bar{\zeta} (x_3+x-2)+x_B)-x\bar{\zeta}\nonumber\\
&&+4 r r_c)\phi^s +(r^2-1)(r^2 (\bar{\zeta}(x-x_3)-x_B)-x \bar{\zeta}) \phi^t ]\big]E_n(t_c)h_c(x_B,x,x_3,b_B,b_3)\nonumber\\
&&+\big[x(r^2-1)[(r^2-\bar{\zeta})\phi^0 +\sqrt{\zeta} \bar{\zeta} (\phi^s-(r^2-1) \phi^t)]-(x_3 \bar{\zeta}-x_B)[(r^2-\bar{\zeta})\phi^0 \nonumber\\
&&+\sqrt{\zeta} r^2 ((r^2-1) \phi^t+\phi^s)]\big] E_n(t_d)h_d(x_B,x,x_3,b_B,b_3) \big\},
\label{eq:m01}
\end{eqnarray}
%%----------------------------------------------------------
where the symbol $\bar{\zeta} = 1-\zeta$, the mass ratios $r=m_{D^{(*)}}/m_B$ and $r_c=m_c/m_B$.
The amplitudes from Fig.~\ref{fig2-Feyn} (b) are written as
%%----------------------------------------------------------
\begin{eqnarray}
F_{TD}&=&8\pi C_F m^4_B f_\rho \int dx_B dx_3\int b_B db_B b_3 db_3 \phi_B \phi_D \big\{\big[
(r+1)[r^2-\bar{\zeta}-x_3 \bar{\zeta}(r-1) (2 r-\bar{\zeta})]\big] \nonumber\\
&&\times E_e(t_m)h_m(x_B,x_3,b_3,b_B)S_t(x_3)+\big[(r^2-\bar{\zeta})[2 r (r_c+1)-r^2 \bar{\zeta}-r_c]-\zeta x_B (2 r-\bar{\zeta}) \big] \nonumber\\
&&\times E_e(t_n)h_n(x_B,x_3,b_B,b_3)S_t(x_B) \big\},
\label{eq:f03}
%\end{eqnarray}
\\%%---------------
%\begin{eqnarray}
M_{TD}&=& 16\sqrt{\frac{2}{3}} \pi C_F m_B^4\int d x_B d x d x_3 \int b_B d b_B b d b \phi_B \phi_D \phi^0 \big\{\big[ x_B [\bar{\zeta}^2-\bar{\zeta}r^2+\zeta r]+\bar{\zeta}x_3 r (\zeta r \nonumber\\
&&+(r+1) (r-1)^2)-\zeta (r-1)^2 (r+1)[(r+2) x-2 (r+1)]+\zeta^2 [x-r^2(x-2)-1]  \nonumber\\
&&+(x-1)(r^2-1)^2\big] E_n(t_o) h_o(x_B, x, x_3, b_B, b)+\big[(r-1) (\bar{\zeta}+r) [x_B+(r^2-1) x]\nonumber\\
&&+\bar{\zeta} x_3[(r-1)^2 (r+1)-\zeta] \big] E_n(t_p) h_p(x_B, x, x_3, b_B, b)\big\}.
\label{eq:m03}
\end{eqnarray}
%%----------------------------------------------------------
The amplitudes from Fig.~\ref{fig2-Feyn} (c) the annihilation diagrams are written as
%%----------------------------------------------------------
\begin{eqnarray}
 F_{A\rho}&=&8\pi C_F m^4_B f_B\int dx_3dx\int bdbb_3db_3\phi_D \big\{\big[ 
((2r r_c-1) (r^2-\bar{\zeta})-(r^2-1)^2 x \bar{\zeta})\phi^0+\sqrt{\zeta} \nonumber\\
&&\times[(r^2-1)(r_c (r^2-\bar{\zeta})-2 r(r^2-1) x) \phi^t+(r_c (r^2-\zeta+1)+2 r (x-x r^2-2))\phi^s] \big]\nonumber\\
&&\times E_a(t_e)h_e(x,x_3,b,b_3)S_t(x)+\big[ (r^2-1)[x_3 \bar{\zeta}^2-\zeta (r^2-\bar{\zeta})]\phi^0+2 \sqrt{\zeta}r(x_3 \bar{\zeta}+\zeta\nonumber\\
&&-r^2+1)\phi^s \big]E_a(t_f)h_f(x,x_3,b_3,b) S_t(|\bar{\zeta}x_3+\zeta|)\big\},
\label{eq:f02}
\end{eqnarray}
%%--------------
\begin{eqnarray}
M_{A\rho}&=&-16\sqrt{\frac{2}{3}} \pi C_F m_B^4\int dx_B dx dx_3\int b_Bdb_Bbdb\phi_B\phi_D\big\{\big[
(r^2-1)[r^2 (x_B+x_3-1)+x_B \nonumber\\
&&+x_3] \phi^0+\zeta [r^4 x-(r^2-1) x_B+\zeta ((r^2-1) x_3-x r^2+x+1)-(r^4+r^2-2) x_3\nonumber\\
&&-x-1]\phi^0 +\zeta^{3/2}r(1-x_3)[(r^2-1)\phi^t+\phi^s]+\sqrt{\zeta}r[\phi^s (x_B+r^2 (x-1)+x_3-x+3)\nonumber\\
&&+(r^2-1)(x_B-x r^2+r^2+x_3+x-1)\phi^t] \big] E_n(t_g)h_g(x_B,x,x_3,b,b_B)+\big[(r^2-\bar{\zeta})\nonumber\\
&&\times [r^2 (x_B-x_3-x+1)+\zeta (r^2 (x_3+x-2)-x+1)+x-1]\phi^0 +\sqrt{\zeta}r[(x_B-x_3 \bar{\zeta}\nonumber\\
&&-\zeta+(r^2-1) (1-x))\phi^s+(1-r^2)(x_B-x_3 \bar{\zeta}-\zeta+(r^2-1) (x-1))\phi^t] \big]\nonumber\\
&&\times E_n(t_h)h_h(x_B,x,x_3,b,b_B)\big\}.
\label{eq:m02}
\end{eqnarray}
Where the $T\rho$, $TD$ and $A\rho$ in the subscript of above expressions stand for $B\to \rho$, $B\to D$ 
transitions and the annihilation Feynman diagrams, respectively.  

The longitudinal polarization amplitudes from Fig.~\ref{fig2-Feyn} (a)  for the decays with a vector 
$\bar{D}^{*0}$ or $D^{*-}$ meson in the final state are written as
%%----------------------------------------------------------
\begin{eqnarray}
F_{T\rho,L}&=&8\pi C_F m^4_B f_{D^*}\int dx_B dx\int b_B db_B b db \phi_B \big\{\big[
[\bar{\zeta}+\bar{\zeta}x (r^2-1)^2 +(2 \zeta -1)r^2]\phi^0\nonumber\\
&&+\sqrt{\zeta} [(1-r^2) (2\bar{\zeta} x (r^2-1)+\bar{\zeta}+r^2)\phi^t+(2\bar{\zeta}x (r^2-1)+\bar{\zeta}-r^2)\phi^s]\big]
            \nonumber\\
&&\times E_e(t_a)h_a(x_B,x,b,b_B)S_t(x)+\big[(r^2-1) [r^2 x_B+\zeta ^2-\zeta (r^2+1)]\phi^0\nonumber\\
&&-2 \sqrt{\zeta} [r^2 (1-x_B)-\bar{\zeta}]\phi^s \big] E_e(t_b)h_b(x_B,x,b_B,b)S_t(|x_B-\zeta|)\big\},
\label{eq:f11}
%\end{eqnarray}
\\%%--------------
%\begin{eqnarray}
M_{T\rho,L}&=&16\sqrt{\frac{2}{3}} \pi C_F m_B^4 \int dx_B dx dx_3 \int b_B db_B b_3 db_3 \phi_B \phi_{D^*}\big\{\big[[rr_c (1-\bar{\zeta}r^2-\zeta ^2)-(r^2-\bar{\zeta}) \nonumber\\
&&\times(\bar{\zeta} x_ 3 (r^2-1)+ x_B(r^2-1)+(\zeta  x-1)r^2 -\zeta (x+1)+1)]\phi^0-\sqrt{\zeta}[(r^2 (x_3 \bar{\zeta} -\bar{\zeta}x \nonumber\\
&&+x_B)-\zeta x+x)\phi^s +(r^2-1)(\bar{\zeta}x (1-r^2) -r^2 ((x_3-2) \bar{\zeta}+x_B))\phi^t ]\big]  \nonumber\\
&&\times E_n(t_c)h_c(x_B,x,x_3,b_B,b_3)+\big[x_B [(\bar{\zeta}+(2 \zeta -1) r^2)\phi^0 +\sqrt{\zeta} r^2 ((r^2-1) \phi^t+\phi^s)]\nonumber\\
&&-\bar{\zeta} x_ 3 [(\bar{\zeta}+(2 \zeta -1) r^2)\phi^0 +\sqrt{\zeta} r^2 ((r^2-1) \phi^t+\phi^s)]+x(r^2-1)[ (\bar{\zeta}+(2 \zeta -1) r^2)\phi^0  \nonumber\\
&&-\sqrt{\zeta} \bar{\zeta} (\phi^s-(r^2-1)\phi^t)]\big] E_n(t_d)h_d(x_B,x,x_3,b_B,b_3)\big\}.
\label{eq:m11}
\end{eqnarray}
%%----------------------------------------------------------
The longitudinal polarization amplitudes from Fig.~\ref{fig2-Feyn} (b) are 
\begin{eqnarray}
F_{TD^{*},L}&=&8\pi C_F m^4_B f_\rho\int dx_B dx_3\int b_B db_B b_3 db_3\phi_B\phi_{D^*} \big\{\big[
\bar{\zeta}+(2 r-1) (r^2-1) x_3 \bar{\zeta}^2+r \nonumber\\
&&\times [\zeta(r^2+2r-\zeta)-r^2-r+1] \big]E_e(t_m)h_m(x_B,x_3,b_3,b_B)S_t(x_3)+\big[r^2[r_c(2 \zeta-1 ) \nonumber\\
&&-\zeta ^2+1]-\bar{\zeta} (\zeta x_B-r_c+r^4)\big] E_e(t_n)h_n(x_B,x_3,b_B,b_3)S_t(x_B)\big\},
\label{eq:f13}
\end{eqnarray}
%%---------------
\begin{eqnarray}
M_{TD^{*},L} &=& -16\sqrt{\frac{2}{3}} \pi C_F m_B^4\int d x_B d x d x_3 \int b_B d b_B b d b \phi_B \phi_{D^*} \phi^0 \big\{\big[ \bar{\zeta} x_B (1-r)(\bar{\zeta}+r)-\bar{\zeta}x_3 r\nonumber\\
&& \times(r^3-\bar{\zeta}(r^2+r-1))-\zeta ^2 (2 r^3-x(r+1)(r-1)^2-2 r+1)+ (x-1)(r^2-1)^2\nonumber\\
&& -\zeta (r+1) (r-1)^2(r x+2x-2)\big]E_n(t_o)h_o(x_B,x,x_3,b_B,b)+\big[ \bar{\zeta} x_3[r^2 (r \bar{\zeta}+2 \zeta -1) \nonumber\\
&& +\bar{\zeta}-r \bar{\zeta}] - (x_B+(r^2-1) x)[\bar{\zeta}-\bar{\zeta}\zeta r+(2 \zeta -1) r^2]\big] E_n(t_p)h_p(x_B,x,x_3,b_B,b)\big\}.\nonumber\\
\label{eq:m13}
\end{eqnarray}
%%----------------------------------------------------------
The longitudinal polarization amplitudes from Fig.~\ref{fig2-Feyn} (c) are 
\begin{eqnarray}
 F_{A\rho,L}&=&-8\pi C_F m^4_B f_B\int dx_3dx\int bdbb_3db_3\phi_{D^*} \big\{\big[
 \sqrt{\zeta} r_c [(r^4-\zeta r^2-\bar{\zeta})\phi^t+ (r^2-\bar{\zeta})\phi^s] \nonumber\\
&& +[\bar{\zeta}(1-x(r^2-1)^2 )+ r^2(2 \zeta -1)]\phi^0\big] E_a(t_e)h_e(x,x_3,b,b_3)S_t(x)+\big[2r\sqrt{\zeta}\bar{\zeta}
\nonumber\\
&&\times ((x_3-1) \bar{\zeta}+r^2) \phi^s+(r^2-1) [\zeta (r^2+\bar{\zeta}(1-x_3)-x_3)+x_3]\phi^0\big]\nonumber\\
&&\times E_a(t_f)h_f(x,x_3,b_3,b)S_t(|\bar{\zeta}x_3+\zeta|)\big\},
\label{eq:f12}
\end{eqnarray}
%%--------------
\begin{eqnarray}
M_{A\rho,L}&=&16\sqrt{\frac{2}{3}} \pi C_F m_B^4\int dx_B dx dx_3\int b_Bdb_Bbdb\phi_B\phi_{D^*} \big\{\big[
 -(x_3\bar{\zeta}+x_B) [(r^2-1) (r^2-\bar{\zeta})\phi^0 \nonumber\\
&& -\sqrt{\zeta}\bar{\zeta} r (r^2-1) \phi^t-\sqrt{\zeta}\bar{\zeta} r \phi^s]-\zeta \bar{\zeta} (x+1)\phi^0+\sqrt{\zeta} \bar{\zeta} r^5 (x-1) \phi^t+\sqrt{\zeta}\bar{\zeta} r^3 ((\zeta-2 x)\phi^t \nonumber\\
&& -(x-1) \phi^s)+\sqrt{\zeta}\bar{\zeta} r((x-\bar{\zeta})\phi^s +(x+\bar{\zeta})\phi^t)-r^4(\zeta x-1) \phi^0 -r^2 (\zeta x(\zeta-2)+1)\phi^0 \big] \nonumber\\
&&\times E_n(t_g)h_g(x_B,x,x_3,b,b_B)-\big[(r^2-\bar{\zeta}) (r^2 (x_B-x_3 \bar{\zeta})+r^2 ( \bar{\zeta}x-1)- \bar{\zeta}(x-1))\phi^0   \nonumber\\
&& -\sqrt{\zeta}\bar{\zeta} r [((x_3-1)\bar{\zeta}+(1-x)r^2+x-x_B)\phi^s-(r^2-1)(x_3 \bar{\zeta}+\zeta-(1-x)r^2-x\nonumber\\
&& -x_B+1) \phi^t ]\big]E_n(t_h)h_h(x_B,x,x_3,b,b_B)\big\}.
\label{eq:m12}
\end{eqnarray}

%%----------------------------------------------------------
The normal along with transverse polarization amplitudes from Fig.~\ref{fig2-Feyn} for the decays with a vector 
$\bar{D}^{*0}$ or $D^{*-}$ are written as
%%----------------------------------------------------------
\begin{eqnarray}
F_{T\rho,{\parallel\perp}}&=&8\pi C_F m^4_B f_{D^*} r \int dx_B dx\int b_B db_B b db \phi_B \big\{\big[ \epsilon _T^{D^*}\cdot \epsilon _T^{\rho} [\sqrt{\zeta} (x (r^2-1) (\phi ^a-\phi ^v)+2 \phi ^v)\nonumber\\
&&+\zeta (2 x (r^2-1)+1)\phi ^T +(1-r^2) \phi ^T]-i \epsilon ^{n v \epsilon _T^{D^*}\epsilon _T^{\rho}} [\sqrt{\zeta} ((x (r^2-1)-2)\phi ^a\nonumber\\
&&-x (r^2-1)\phi ^v)+\zeta(2 x (r^2-1)+1)\phi ^T+(r^2-1)\phi ^T]\big]E_e(t_a)h_a(x_B,x,b,b_B)S_t(x)\nonumber\\
&&+\sqrt{\zeta}\big[\epsilon _T^{D^*}\cdot \epsilon _T^{\rho}[(\zeta-x_B-r^2+1)\phi ^v +(\bar{\zeta}+x_B-r^2)\phi ^a]+i \epsilon ^{n v \epsilon _T^{D^*}\epsilon _T^{\rho}} [ (\zeta-x_B -r^2\nonumber\\
&&+1)\phi ^a-(\zeta-x_B +r^2-1)\phi ^v]\big] E_e(t_b)h_b(x_B,x,b_B,b)S_t(|x_B-\zeta|)\big\},
\label{eq:f21}
%\end{eqnarray}
\\%%--------------
%\begin{eqnarray}
M_{T\rho,{\parallel\perp}}&=&16\sqrt{\frac{2}{3}} \pi C_F m_B^4 \int dx_B dx dx_3 \int b_B db_B b_3 db_3 \phi_B \phi_{D^*}\big\{\big[ \epsilon _T^{D^*}\cdot \epsilon _T^{\rho}[\zeta ^{3/2} r_c (\phi ^a-\phi ^v) \nonumber\\
&& +\sqrt{\zeta} r_c((r^2-1) \phi ^a+(r^2+1)\phi ^v)+r (r^2-1) (x_B+x_3-1)\phi ^T-\zeta r ((r^2-1) \nonumber\\
&& \times (x_3+x)-2r^2+1)\phi ^T]-i \epsilon ^{n v \epsilon _T^{D^*}\epsilon _T^{\rho}} [\zeta ^{3/2} r_c (\phi ^a-\phi ^v)-\sqrt{\zeta} r_c ((r^2+1) \phi ^a\nonumber\\
&& +(r^2-1) \phi ^v)-r (r^2-1) (x_B+x_3-1)\phi ^T+\zeta r ((x_3-x)(r^2-1)+1)\phi ^T]\big] \nonumber\\
&& \times E_n(t_c)h_c(x_B,x,x_3,b_B,b_3)+r\big[\epsilon _T^{D^*}\cdot \epsilon _T^{\rho}[2 \sqrt{\zeta} (x_B+x (r^2-1)-x_3 \bar{\zeta}) \phi ^v\nonumber\\
&&+(r^2-1) (x_B-x\zeta-x_3 \bar{\zeta})\phi ^T]+i \epsilon ^{n v \epsilon _T^{D^*}\epsilon _T^{\rho}} [2 \sqrt{\zeta} (x_B+x (r^2-1)-x_3 \bar{\zeta}) \phi ^a \nonumber\\
&&+(r^2-1) (x_B+x\zeta -x_3 \bar{\zeta})\phi ^T ]\big]E_n(t_d)h_d(x_B,x,x_3,b_B,b_3)\big\},
\label{eq:m21}
\end{eqnarray}
%%----------------------------------------------------------
\begin{eqnarray}
F_{TD^{*},{\parallel\perp}}&=&8\pi C_F m^4_B f_\rho\sqrt{\zeta}\int dx_B dx_3\int b_B db_B b_3 db_3\phi_B\phi_{D^*} \big\{\big[ 
\epsilon _T^{D^*}\cdot \epsilon _T^{\rho}[x (r^2-1) (2\bar{\zeta}-r)\nonumber\\
&& +\bar{\zeta}+r^2+2r]-i \epsilon ^{n v \epsilon _T^{D^*}\epsilon _T^{\rho}} [x (r^2-1) (r-2\bar{\zeta})-\bar{\zeta}+r^2]\big]E_e(t_m)h_m(x_B,x_3,b_3,b_B) \nonumber\\
&& \times S_t(x_3)+r\big[\epsilon _T^{D^*}\cdot \epsilon _T^{\rho}[\zeta-x_B+2 r_c -r^2+1]-i \epsilon ^{n v \epsilon _T^{D^*}\epsilon _T^{\rho}} [\bar{\zeta}+x_B-r^2]\big] \nonumber\\
&& \times E_e(t_n)h_n(x_B,x_3,b_B,b_3)S_t(x_B)\big\},
\label{eq:f23}
\end{eqnarray}
%\\%%---------------
\begin{eqnarray}
M_{TD^{*},{\parallel\perp}}&=&16\sqrt{\frac{2}{3}} \pi C_F m_B^4\sqrt{\zeta}\int dx_B dx dx_3\int b_Bdb_Bbdb\phi_B\phi_{D^*} \big\{\big[ \epsilon _T^{D^*}\cdot \epsilon _T^{\rho}[r^2 (\bar{\zeta} ((2-x) \phi ^v\nonumber\\
&& -x \phi ^a)+(\bar{\zeta} x_3+x_B)(\phi ^a-\phi ^v))+ \bar{\zeta} x(\phi ^a+\phi ^v)]-i \epsilon ^{n v \epsilon _T^{D^*}\epsilon _T^{\rho}} [r^2 ((\bar{\zeta}(x-x_3)\nonumber\\
&& -x_B)\phi ^v +(\bar{\zeta}(x_3+x-2)+x_B)\phi ^a )-\bar{\zeta} x (\phi ^a+\phi ^v)] \big] E_n(t_o)h_o(x_B,x,x_3,b_B,b)\nonumber\\
&&+\big[ \epsilon _T^{D^*}\cdot \epsilon _T^{\rho}[ (r^2 (x_B-x_3 \bar{\zeta})-x \bar{\zeta}(r^2-1))\phi ^a+(x (r^2-1) (2 r-\bar{\zeta})-(r-2)\nonumber\\
&&\times r (x_B-x_3 \bar{\zeta}))\phi ^v ]+i \epsilon ^{n v \epsilon _T^{D^*}\epsilon _T^{\rho}} [(x (r^2-1) (2 r-\bar{\zeta})- r(r-2) (x_B-x_3 \bar{\zeta}))\phi ^a\nonumber\\
&& + (r^2 (x_B-x_3 \bar{\zeta})-\bar{\zeta}x (r^2-1))\phi ^v]\big] E_n(t_p)h_p(x_B,x,x_3,b_B,b)\big\},
\label{eq:m23}
\end{eqnarray}
%%----------------------------------------------------------
\begin{eqnarray}
 F_{A\rho,{\parallel\perp}}&=&8\pi C_F m^4_B f_B r\int dx_3dx\int bdbb_3db_3\phi_{D^*}\big\{\big[ 
\epsilon _T^{D^*}\cdot \epsilon _T^{\rho}[\sqrt{\zeta} (x (r^2-1) (\phi ^a-\phi ^v)-2 \phi ^v)\nonumber\\
&& -r_c(r^2-\zeta-1) \phi ^T]+i\epsilon ^{n v \epsilon _T^{D^*}\epsilon _T^{\rho}} [\sqrt{\zeta} (x(r^2-1)\phi ^v- (x(r^2-1)+2)\phi ^a) +(r^2-\bar{\zeta}) \nonumber\\
&& \times r_c \phi ^T] \big]E_a(t_e)h_e(x,x_3,b,b_3)S_t(x)+\sqrt{\zeta}\big[\epsilon _T^{D^*}\cdot \epsilon _T^{\rho} 
[(\bar{\zeta}x_3+\zeta-r^2 + 1)\phi ^v+(\bar{\zeta}x_3 \nonumber\\
&& +\zeta+r^2-1)\phi ^a]+i \epsilon ^{n v \epsilon _T^{D^*}\epsilon _T^{\rho}} [(\bar{\zeta}x_3+\zeta+r^2-1)\phi ^v+(\bar{\zeta}x_3+\zeta-r^2+1)\phi ^a]\big] \nonumber\\
&& \times E_a(t_f)h_f(x,x_3,b_3,b)S_t(|\bar{\zeta}x_3+\zeta|)\big\},
\label{eq:f22}
\end{eqnarray}
%%--------------
\begin{eqnarray}
M_{A\rho,{\parallel\perp}}&=&16\sqrt{\frac{2}{3}} \pi C_F m_B^4\int dx_B dx dx_3\int b_Bdb_Bbdb\phi_B\phi_{D^*} \big\{\big[
\epsilon _T^{D^*}\cdot \epsilon _T^{\rho}[ (r^2 x_B(r^2-1)\nonumber\\
&& +\bar{\zeta}r^2((r^2-1)(x_3-1)+\zeta ) -\bar{\zeta}\zeta x (r^2-1) )\phi ^T-2 \sqrt{\zeta} r \phi ^v]+i \epsilon ^{n v \epsilon _T^{D^*}\epsilon _T^{\rho}}[(r^2 (\bar{\zeta}r^2\nonumber\\
&& -\bar{\zeta}^2-(\bar{\zeta}x_3+x_B)(r^2-1))-\bar{\zeta}\zeta x (r^2-1))\phi ^T-2 \sqrt{\zeta} r \phi ^a] \big] E_n(t_g)h_g(x_B,x,x_3,b,b_B)\nonumber\\
&& +(r^2-1) \big[\epsilon _T^{D^*}\cdot \epsilon _T^{\rho}[r^2 (x_B-x_3)+\zeta (r^2 (x_3-1)+x-1)+\zeta ^2 (1-x)]-i \epsilon ^{n v \epsilon _T^{D^*}\epsilon _T^{\rho}} \nonumber\\
&& \times [r^2 (x_B-x_3)+\zeta (r^2 (x_3-1)-\bar{\zeta}(x-1))]\big]\phi ^T E_n(t_h)h_h(x_B,x,x_3,b,b_B)\big\}.
\label{eq:m22}
\end{eqnarray}
%%----------------------------------------------------------
The hard functions $h_i$, the hard scales $t_i$ with $i \in\{a,b,c,d,e,f,g,h,m,n,o,p\}$, and the evolution factors $E_{e,a,n}$ 
have their explicit expression in the Appendix of Ref.~\cite{2311.00413}. 

%%<<<<><><><><><><><><><><><><><><><><><><><><><><><><><><><><><><><><>>>>%%
\section{Decay amplitude for $B^+\to\bar{D}^{0} [a_0(980,1450)^+\to] K^+\bar{K}^0$}  
\label{sec-appx}

The decay constants of the pseudoscalar meson $P$ and the scalar meson $S$ are defined by
\begin{eqnarray}
 \label{eq:decayc}
     &\langle P(p)|\bar q_2\gamma_\mu\gamma_5 q_1|0\rangle=-if_Pp_\mu,   \nonumber\\
     &\langle S(p)|\bar q_2\gamma_\mu q_1|0\rangle=f_S p_\mu,               \nonumber\\
    &\langle S|\bar q_2q_1|0\rangle=m_S\bar f_S.                                           \nonumber
\end{eqnarray}

When the kaon pair $K^+\bar{K}^0$ originated from the intermediate $a_0(980,1450)^+$, we have~\cite{jhep2003-162}
\begin{eqnarray}
 \label{aKK}
  \langle K^+\bar{K}^0 | \bar{d} u|0  \rangle &=&  \langle K^+\bar{K}^0 | a_0  \rangle \frac{1}{\mathcal{D}_{a_0}} 
         \langle  a_0  | \bar{d} u|0  \rangle                                                                         %  \nonumber \\
       % &=&
         \frac{g_{a_0KK}}{\mathcal{D}_{a_0}} \langle  a_0  | \bar{d} u|0  \rangle,    
\end{eqnarray}
with the denominator~\cite{prd95.032002,prd102.012002}
\begin{eqnarray}
   {\mathcal{D}_{a_0}}={m_{a_0}^2-s-i(g^2_{\pi\eta}\rho_{\pi\eta}
                 +g^2_{\bar{K}^0K}\rho_{\bar{K}^0K}+g^2_{\pi\eta'}\rho_{\pi\eta'})}. \quad %\nonumber  
\end{eqnarray}
The $B \to D^{(*)}$ matrix element is described by the transition form factors~\cite{zpc29-637}
\begin{eqnarray}
    \label{B2D}
    && \langle D(p^\prime) | \bar c \gamma^\mu b | {B}(p)\rangle  =
                  F^{BD}_0(q^2)\, \frac{m_B^2-m_D^2}{q^2}\, q^\mu  \;\;                               \nonumber\\
    && \phantom{\langle D(p^\prime) | \bar c} 
                  + F^{BD}_1(q^2) \big[(p+p^\prime)^\mu - \frac{m_B^2-m_D^2}{q^2}\,  q^\mu \big]\!, \quad    
\end{eqnarray}
where  $q=p-p^\prime$.  
We parameterize the matrix element for the $B \to a_0$ transition in terms of form factors $F_0^{Ba}$ and $F_1^{Ba}$ as
\begin{eqnarray}
   \label{B2a0}
    &&\langle a_0(p^\prime) | \bar q \gamma^\mu  \gamma_5 b | {B}(p)\rangle  
       = i F^{Ba}_0(q^2)\, \frac{m_B^2-m_{a_0}^2}{q^2}\, q^\mu                                               \nonumber\\
    && \phantom{\langle a_0(p^\prime) | \bar u }
         + i F^{Ba}_1(q^2) \big[(p+p^\prime)^\mu - \frac{m_B^2-m_{a_0}^2}{q^2}\, q^\mu \big]\!. \qquad  
\end{eqnarray}

With Eqs.~(\ref{eq_das_ch13})-(\ref{eq-das_ch24}) and related transition form factors above, we have the decay 
amplitude
\begin{eqnarray}
    &&\mathcal{M}(B^+\to\bar{D}^{0} [a_0^+\to] K^+\bar{K}^0)
                         =\frac{G_F}{\sqrt2}V^*_{cb}V_{ud}                                                                                            \nonumber\\
    && \phantom{ \mathcal{M}_S(B^+\to} \times\big[ a_1(m^2_B-m^2_{a_0})f_D F_0^{Ba} (m^2_D)             %      \nonumber\\
    %&& \phantom{ \mathcal{M}_S(B^+\to} 
    + a_2(m^2_B-m^2_{D})f_{a_0} F_0^{BD}(s) \big] 
               \frac{g_{a_0KK}}{\mathcal{D}_{a_0}}\!.\qquad          \label{da_Bua0}                
\end{eqnarray}               
The expressions and related parameters for $F_0^{Ba}$ and $F_0^{BD}(s)$ are found in 
Refs.~\cite{prd87.114001,prd92.094016,jhep0106-067,prd69.074025}.

%=============== Refs ===============%  

\end{document}